\documentclass[11pt]{article}
\usepackage{amssymb,geometry}
\usepackage{graphicx}
\usepackage[sc,small,margin=1cm]{caption}
\newcommand{\eqref}[1]{(\ref{#1})}
\usepackage{hyperref}
\hypersetup{
    colorlinks=true,
    linkcolor=black,
    citecolor=black,
    filecolor=black,
    urlcolor=blue,
}

\newcommand{\be}{\begin{equation}}
\newcommand{\ee}{\end{equation}}

\begin{document}
\begin{center}
{\bf {\Large  }}
\end{center}

\begin{center}

{\LARGE \bf \sc  Massive Theta Lifts}\\[5mm]

\vspace{6mm}
\normalsize
{\large  Marcus Berg${}^{1}$ and Daniel Persson${}^2$}

\vspace{10mm}
${}^{2}${\it Department of Physics\\
Karlstad University, 651 88 Karlstad, Sweden}\\[2mm]
${}^{2}${\it Department of Mathematical Sciences\\
Chalmers University of Technology \\ University of Gothenburg \\ 41296 Gothenburg, Sweden}

\vspace{20mm}

\hrule

\vspace{5mm}

 \begin{tabular}{p{14cm}}

We use Poincar\'e series 
for massive Maass-Jacobi forms  
to define a ``massive theta lift'', and apply it
to the examples of the constant function and the modular invariant $j$-function, with the Siegel-Narain theta function as integration kernel. 
These theta integrals are deformations of known one-loop string threshold corrections.
Our massive theta lifts fall off exponentially,
so some Rankin-Selberg integrals are finite without Zagier renormalization. 
\end{tabular}

\vspace{6mm}
\hrule
\end{center}

\thispagestyle{empty}

\newpage
\setcounter{page}{1}

\setcounter{tocdepth}{2}
\tableofcontents


\bigskip

\section{Introduction}

The theory of automorphic forms is a wide-ranging area of mathematics, as demonstrated, for example, through Wiles's proof of Fermat's last theorem. This is part of a ``grand unified theory of mathematics'' called the \emph{Langlands program}, which  relates automorphic forms to group representation theory, geometry and number theory in deep and unexpected ways. Many interesting properties of automorphic forms are captured by their Fourier coefficients. Classically, these coefficients often carry a wealth of interesting information such as eigenvalues of Hecke operators and counts of rational points of elliptic curves. For higher-rank groups, such Fourier coefficients  play a crucial role in analyzing automorphic $L$-functions by the Langlands--Shahidi method (see, e.g., \cite{Shahidi}).

The moduli space of toroidal string compactifications is a symmetric space $G(\mathbb{R})/K$.
String scattering amplitudes that preserve a maximal amount of supersymmetry
are functions on this space and they have to be covariant (equivariant) under a discrete subgroup $G(\mathbb{Z})\subset G(\mathbb{R})$ (U-duality group). Physical considerations reveal that these functions are highly constrained: (i) they must be eigenfunctions of the Laplacian on $G(\mathbb{R})/K$ with specific eigenvalues; (ii) they must have a small number of constant terms with specific growth conditions, and (iii) their Fourier coefficients are supported only on a subspace of all possible `instanton charges'. Mathematically this means that these functions are automorphic forms attached to  certain small automorphic representations.  The vanishing of the associated Fourier coefficients precisely matches with the expected physical properties of quartic gravitational scattering amplitudes in  string compactifications.  

The focus of the present work is to study  \emph{massive theta lifts}. Theta lifts are given by certain integrals of automorphic forms, producing new automorphic objects. They have a representation theoretic origin, knowns generally as \emph{theta correspondences}, in which an automorphic form $\varphi$ attached to a group $G$ is transferred to an automorphic form $\varphi^{\prime}$ on another group $G^\prime$. If $G\subset G^{\prime}$ we call this a \emph{lifting} of the automorphic form. Practically, a theta lift is achieved by integrating the automorphic form $\varphi$ against a theta function. A theta function is an example of a modular form corresponding to a \emph{small} representation, as mentioned above. Theta lifts play an important role in mathematics, in particular through their explicit realization of the transfer of automorphic representations, known as the \emph{principle of functoriality} in the Langlands program. On the other hand, theta lifts also appear naturally in string theory where they correspond to certain scattering amplitudes. More precisely string S-matrix elements are naturally computed as integrals over worldsheet moduli space, which is the double coset space $SL(2,\mathbb{Z})\backslash SL(2, \mathbb{R})/SO(2)$. Integrating a Siegel-Narain theta function against this moduli space produces an automorphic form on the orthogonal group $SO(d,d)$. The prototypical example of a theta lift of this form is given by 
\begin{equation}
\int_{SL(2,\mathbb{Z})\backslash \mathbb{H}} \Theta_{d,d}(\tau; g) \frac{d^2\tau}{\tau_2^2}, \qquad g\in SO(d,d),
\end{equation}
where $\Theta_{d,d}(\tau;g)$ is a Siegel-Narain theta function on $\mathbb{H}\times SO(d,d)/(SO(d)\times SO(d))$, with $\mathbb{H}$ being the standard upper half plane.  This is a theta lift of the constant function $1$, i.e. the trivial representation, and produces an automorphic form on $SO(d,d)$ attached to the minimal representation. If we insert a non-trivial modular form $f(\tau)$ we can obtain a wider class of representations. 

The purpose of this paper is to analyze a ``massive deformation'' of the theta lift. Massive modular forms were introduced in \cite{Bergman:2002hv,Berg:2019jhh}, with motivation from string theory. From a mathematical point of view, this is a one-parameter family of deformations of modular forms. The parameter $\mu$ can be interpreted as a ``mass'' in a string theory context. We will consider theta lifts of these massive deformations. Rather than developing a general theory, our aim is to work out a few examples in detail. It is interesting to inquire about the role of representation theory here. We offer some comments on this question in the conclusions.

This paper is organized as follows. In section \ref{MassivePrelims} we introduce some necessary background on modular forms (Jacobi forms and Maass forms), after which we introduce their massive deformations. We then consider the explicit Fourier expansion of the massive Eisenstein series. This result is needed later for the massive theta lifts. Section \ref{ThetaRSZ} introduces the Rankin-Selberg-Zagier transform, which is used to compute evaluate standard theta lifts. We consider the explicit examples of lifting both the trivial function 1 as well as the modular invariant $j$-function. The massive theta lift is introduced in section \ref{massivethetalift}. Here we consider various approaches to computing the massive theta lift of the constant function. The following section \ref{liftingjfunction} then treats the case of the massive theta lift of the $j$-function. We end in section \ref{conclusions} with some conclusions and suggestions for future work. The paper also contains several appendices with  calculational details and various digressions.

\section{Massive Maass-Jacobi forms}
\label{MassivePrelims}
In this section we first introduce the necessary background on modular forms and their massive deformations (sections \ref{prel} and \ref{massiveprel}). In section \ref{FourierMassive} we then present the complete Fourier expansion of the massive non-holomorphic Eisenstein series. This result will be used in subsequent sections when we compute the massive theta lift.  
\subsection{Some preliminaries}
\label{prel}
A {\it modular form} of weight $k$ is a holomorphic function $f\, : \mathbb{H}\to \mathbb{C}$ on the complex upper half-plane $\mathbb{H}$, that transforms according to $f\left(\frac{a\tau+b}{c\tau+d}\right)=(c\tau+d)^k f(\tau)$ with respect to modular transformations.
 Modular forms admit a Fourier expansion  $f(\tau)=\sum_{n\geq 0} c(n)q^n$, where $q=e^{2\pi i \tau}$. In this context the Fourier expansion is also referred to as a $q$-{\it expansion} since it can be written as a power series in $q$. As we will see below, this is not the case when we relax the condition of holomorphy, which will be crucial for the project. The $n=0$ part of the Fourier sum is called the \emph{constant term}. 

As already alluded to, one can also consider modular forms that are not holomorphic; these are known as \emph{Maass forms}. A classical example is the non-holomorphic Eisenstein series
\begin{equation}
E_s(\tau)=\frac{1}{2}\sum_{\gcd(m,n)=1} \frac{y^{s}}{|m+n\tau|^{2s}},\qquad \tau=x+iy\in \mathbb{H}.
\label{nonholo}
\end{equation}
This function is invariant under $SL(2,\mathbb{Z})$ (i.e.~of weight 0), and converges absolutely for $\Re(s)>1$, but can be analytically continued to a meromorphic function of all $s\in \mathbb{C}$ away from a simple pole at $s=1$. It is an eigenfunction of the Laplacian on $\mathbb{H}$ with eigenvalue $s(s-1)$. The non-holomorphic Eisenstein is a prototype of the kind of functions that are relevant for this paper. Let us therefore analyze its Fourier coefficients in some detail. Since it is a non-holomorphic function, the Fourier expansion is not a simple $q$-expansion.

The Fourier coefficients $c(m;\tau)$ of $E_s(\tau)$ now depend on $\tau=\tau_1+i\tau_2$ and are indexed by $m\in \mathbb{Z}$. The non-constant coefficients $(m\neq 0)$ are given by 
\begin{equation}
c(m; \tau)=\int_0^1 E_s(\tau+u) e^{-2\pi i m u} du={2 \over \xi(2s)}\tau_2^{1/2}|m|^{s-1/2}\sigma_{1-2s}(m)K_{s-1/2}(2\pi|m|\tau_2)e^{2\pi i \tau_1},
\label{sl2coeff}
\end{equation}
where $K_t$ is the modified Bessel function, $\xi(t)$  is the completed Riemann zeta function and $\sigma_t(m)=\sum_{d|m} d^{t}$ is the  sum over divisors of $m$. The $K$-Bessel function appears as a solution to the Laplacian equation that respects the moderate-growth condition.

The non-holomorphic Eisenstein series discussed above is an example of a \emph{Maass form}. A Maass form is a smooth, $\mathbb{C}$-valued function on the upper half plane $\mathbb{H}$, with prescribed transformation properties with respect to $SL(2,\mathbb{Z})$ and at most polynomial growth at the cusps. It is furthermore required to be an eigenfunction with respect to certain differential operators on $\mathbb{H}$. 

A {\it Jacobi form of weight $k$ and index $m$} is a function $\phi : \mathbb{H}\times \mathbb{C}\to \mathbb{C}$ that transforms as follows with respect to $SL(2,\mathbb{Z})\ltimes \mathbb{Z}^2$:
\begin{equation} \phi\left(\frac{a\tau+b}{c\tau+d}, \frac{z+\lambda\tau+\nu}{c\tau+d}\right)=\exp\left[2\pi i \left(-c\frac{(z+\lambda\tau+\mu)^2}{c\tau+d}+\lambda^2\tau+2\lambda z\right)\right](c\tau+d)^k\phi(\tau, z).
\end{equation}
Examples include the Jacobi theta functions.

\subsection{Massive deformations}
\label{massiveprel}
Following \cite{Berg:2019jhh}, a {\it massive Maass form of weight $k$} is, for each $\mu\in \mathbb{R}_{\geq 0}$, a function $f_\mu : \mathbb{H}\to \mathbb{C}$ that transforms as a Maass form of weight $k$ with respect to $SL(2, \mathbb{Z})$, and satisfies the differential equation 
\begin{equation} \label{Delta}
\Delta_{\tau, k}f_\mu(\tau)= \Big(g_2(\mu)\partial_\mu^2+g_1(\mu)\partial_\mu + g_0(\mu)\Big)f_\mu(\tau),
\end{equation}
for certain functions $g_j: \mathbb{R}_{\geq 0}\to \mathbb{C}$, $j=0,1,2$. In general,  $f_\mu(\tau)$ should also satisfy a certain higher order differential equation, but we focus on the property \eqref{Delta} of $f_\mu(\tau)$. See \cite{Berg:2019jhh} for more details. We say that $f_\mu(\tau)$ is a massive deformation of the weight $k$ Maass form $f(\tau):=\lim_{\mu \to 0+}f_{\mu}(\tau)$.

In order to introduce massive Maass-Jacobi forms we consider the following differential operator
\begin{equation}
\Delta_{z, k, m}:= 2\tau_2\partial_z \partial_{\bar{z}}+8\pi i \tau_2 m\partial_{\bar{z}}-2\pi i m.
\end{equation}
A {\it massive Maass-Jacobi form of weight $k$ and index $m$} is, for each $\mu\in \mathbb{R}_{\geq 0}$, a function $\phi_\mu : \mathbb{H}\times \mathbb{C}\to \mathbb{C}$, that transforms like a Jacobi form of weight $k$ and index $m$ and furthermore satisfies the condition
\begin{equation}
\Delta_{z, k, m}\phi_\mu(\tau, z)=-\Big(G_2(\mu)\partial_\mu^2+G_1(\mu)\partial_\mu+G_0(\mu)\Big)\phi_\mu(\tau, z),
\end{equation}
for certain functions $G_j:\mathbb{R}_{\geq 0}\to \mathbb{C}, j=0,1,2$.

As an example, consider the following function
\begin{equation}
\mathcal{E}_{s, \mu}(\tau):={2\pi^s \over \Gamma(s)} (\mu\tau_2)^{s/2}\sum^{\quad \prime}_{(m,n)=1}\left(\frac{\sqrt{\mu\tau_2}}{|m\tau+n|}\right)^s K_s\left(2\pi \sqrt{\frac{\mu}{\tau_2}}|m\tau+n|\right).
 \label{MJ0}
\end{equation}
The prime on the sum indicates that the term $(m,n)= (0,0)$ is absent, and we only sum over coprime $m$ and $n$. This is trivially a massive Maass-Jacobi form of weight $0$ and and index $0$ (it has no elliptic parameter $z$), and in particular \eqref{MJ0} is a massive Maass form. It will be our main test case, but we believe that the calculations below can be of interest also for more general classes of massive Maass-Jacobi forms than  \eqref{MJ0}. 

In Appendix \ref{holo} we also discuss massive deformations of some holomorphic modular forms. 

\subsection{Fourier expansion of massive Eisenstein series}
\label{FourierMassive}
It is convenient to rewrite the massive Eisenstein series as a Poincar\'e series as follows: 
\begin{equation} \label{Poinc0}
\mathcal{E}_{s, \mu}(\tau, z)=\sum_{\gamma\in \Gamma_{\infty}\backslash SL(2,\mathbb{Z})}\sigma_{\mu}(\gamma \cdot \tau_2),
\end{equation}
where we have defined the {\it massive seed function}
\be
\sigma_{\mu}(\tau_2):={2\pi^s \over \Gamma(s)} (\mu\tau_2)^{s/2}K_s(2\pi \sqrt{\mu/\tau_2}).
\ee
Just like for ordinary non-holomorphic Eisenstein series the seed function has no $\tau_1$ dependence, which simplifies the computation of the Fourier modes. 

The constant term (in $\tau_1$) is the sum of the seed $\sigma_{\mu}(\tau_2)$ and the $n=0$ Fourier mode:
\begin{eqnarray*}
f_{\mu,0}(\tau_2)&=&\sigma_{\mu}(\tau_2) + {\pi^s \over \Gamma(s)}\sum_{c>0}S(0,0;c)\int_{\mathbb R} 2\left(\mu \tau_2 \over c^2(\tilde{\omega}^2+\tau_2^2)\right)^{s/2}\! K_s\left(2\pi \sqrt{\mu \over \tau_2/(c^2(\tilde{\omega}^2+\tau_2^2))}\right)d\tilde{\omega}\\
 &=& \sigma_{\mu}(\tau_2)+ {\pi^s \over \Gamma(s)}\sum_{c>0} \varphi(c) c^{-s-1/2}\cdot 2 \tau_2^{3/4-s/2}
\mu^{s/2-1/4}K_{s-1/2}(2\pi c \sqrt{\mu\tau_2 }) \;. 
\end{eqnarray*}
Viewed as a deformation of the usual Eisenstein series at $\mu=0$, we can attempt to series expand in $\mu$. 
Note, however, that this ruins the exponential suppression of the Bessel function at large arguments,
and that even for small $\mu$, there will be some large $c$ or large $\tau_2$ where the series is not a good approximation
to the original expression. 
If we proceed formally anyway, 
the sum  has the following series expansion in $\mu$:
\begin{eqnarray*}
f_{\mu,0}(\tau_2)- \sigma_{\mu}(\tau_2) &=&{\xi(2s-1) \over \xi(2s) } \tau_2^{1-s}
-{\pi^2 \xi(4-2s) \over (s-1)\xi(3-2s)}\mu  \tau_2^{2-s} + {\pi^4 \xi(6-2s) \over 2(s-2)(s-1)\xi(5-2s)}\mu^2  \tau_2^{3-s}
+ \ldots
\end{eqnarray*}
where $\xi(s)= \pi^{-s/2}\Gamma(s/2)\zeta(s)$, and we used ($\varphi$ is the Euler totient function)
\be  \label{totient}
 \sum_{c>0}{S(0,0;c) \over c^{2s}}  =
 \sum_{c>0}{\varphi(c) \over c^{2s}}  = {\zeta(2s-1) \over \zeta(2s)}.
\ee
In particular, there are a priori odd powers of  $c$, that come with odd powers of $\sqrt{\mu}$, but by eq.\ \eqref{totient}, they sum to zero. 
Note that the pole at $s=1$ now receives corrections order by order in $\mu$.  For the $s=1$ case, note that $(s-1)\zeta(3-2s)$ is finite, but all higher terms contribute to the residue.

For the nonzero Fourier modes we find:
\begin{eqnarray*}
f_{\mu,n}(\tau_2)&=&{\pi^s \over \Gamma(s)}  \sum_{c>0}S(0,n;c)\int_{\mathbb R}e^{-2\pi i n \tilde{\omega}}\, 2\left(\mu \tau_2 \over c^2(\tilde{\omega}^2+\tau_2^2)\right)^{s/2}\! K_s\left(2\pi \sqrt{\mu \over \tau_2/(c^2(\tilde{\omega}^2+\tau_2^2))}\right)d\tilde{\omega}\\
&=& {\pi^s \over \Gamma(s)} \sum_{c>0} S(0,n;c) c^{-2s}\cdot 2 \tau_2^{3/4-s/2}
(n^2\tau_2+ \mu c^2)^{s/2-1/4}K_{s-1/2}(2\pi \sqrt{\tau_2}\sqrt{n^2\tau_2+\mu c^2})
\end{eqnarray*}
where we have used the Bessel identity in Appendix \ref{FourierBessel}. Now, if we write each Fourier coefficient as
\be
f_{\mu,n}(\tau_2)=f_{n}^{(0)}(\tau_2) + f_{n}^{(1)}(\tau_2) \mu + f_{n}^{(2)}(\tau_2)\mu^2 +  f_{n}^{(3)}(\tau_2)\mu^3  + \ldots
\ee
we have
\begin{eqnarray} \label{fFourier}
f_{n}^{(0)}(\tau_2)  &=& {2 \over \xi(2s) }\tau_2^{1/2}|n|^{s-1/2} \sigma_{1-2s}(n)K_{s-1/2}(2\pi|n| \tau_2) \\
f_{n}^{(1)}(\tau_2)  &=& 
{\pi  \over (s-1) \xi(3-2s)}  \tau_2^{-1/2}|n|^{s-5/2}\sigma_{3-2s}(n) \left(c_{-}^{(1)}K_{s-1/2}(2|n|\pi\tau_2)
 + c_{+}^{(1)} K_{s+1/2}(2\pi |n| \tau_2)\right) \nonumber \\
 f_{n}^{(2)}(\tau_2)  &=& 
{\pi^2 \over 4(s-1)(s-2) \xi(5-2s)}  \tau_2^{-3/2}|n|^{s-9/2}\sigma_{5-2s}(n) \left(c_{-}^{(2)}K_{s-1/2}(2|n|\pi\tau_2)
 + c_{+}^{(2)} K_{s+1/2}(2\pi |n| \tau_2)\right) \nonumber  \\
f_{n}^{(3)}(\tau_2)  &=&
{\pi^3 \over 24(s-1)(s-2)(s-3)\xi(7-2s)}  \tau_2^{-5/2}|n|^{s-13/2}\sigma_{7-2s}(n) \left(c_{-}^{(1)}K_{s-1/2}(2|n|\pi\tau_2)
 + c_{+}^{(1)} K_{s+1/2}(2\pi |n| \tau_2)\right)  \nonumber   
\end{eqnarray}
where the coefficients of $K_{s-1/2}$ and $K_{s+1/2}$ at each order in $\mu$ are, with the shorthand $x = 2\pi |n|\tau_2$,
\be
\begin{array}{rclrcl}
c_{-}^{(1)} &=& 1-2s&  c_{+}^{(1)}  &=& -x \\
c_{-}^{(2)} &=& 4(s-2)s+x^2+3 & \; , \quad c_{+}^{(2)}  &=& (3-2s)x   \; .  \\
c_{-}^{(3)} &=& (2s-3)(4(s-3)s+2x^2+5) & \; , \quad c_{+}^{(3)}  &=& -x(4(s-4)s+x^2+15)  \; . 
\end{array}
\ee
Here we used that the sum $S(0,n;c)$ produces ($\sigma$ is the divisor function)
\be
 \sum_{c>0}{S(0,n;c) \over c^{2s}}  = {\sigma_{1-2s}(n) \over \zeta(2s)} \; . 
\ee
Both here and for the constant term, we the kept the coefficient $f_{n}^{(0)}(\tau_2)$ of the zeroth order term in its original form, but for the higher-order terms we used the functional relation for $\zeta$, so that the argument of the zeta function increases with order in $\mu$.  

A minor comment is that the order $\mu$ term can be simplified to a single Bessel function $K_{s-3/2}$,  but we have chosen to keep this form to exhibit a pattern that persists to higher order:
the expansion in $\mu$ can be arranged as an expansion of the coefficients of $K_{s-1/2}$ and $K_{s+1/2}$.

The zeroth order (in $\mu$) coefficient $f_{n}^{(0)}(\tau_2)$ vanishes as $s\rightarrow 0$. We remind the reader
that the usual Eisenstein series has an apparently ill-defined Poincar\'e series \eqref{Poinc0} as $s\rightarrow 0$, 
whereas the Fourier expansion in $\tau_1$ gives the value 1 for the ordinary massless Eisenstein series as $s\rightarrow 0$.

%

 The polynomials that are generated for $s\rightarrow 0$ are reverse Bessel polynomials
that are generated by
\be
{1 \over \sqrt{1-2t}}e^{x(1-\sqrt{1-2t})} = \sum_{n=0}^{\infty}\theta_n(x){t^n \over n!} \; . 
\ee
with  $\theta_n(x) =\sqrt{2/\pi}x^{n+1/2}e^x K_{n+1/2}(x)$. 

For the $s\rightarrow 1$ case, note as above that $(s-1)\zeta(3-2s)$ is finite at $s=1$, but all higher terms contribute to the residue. The residue can be subtracted order by order in $\mu$, 
as in the massless case,
that produces a finite remainder with a Fourier series that converges rapidly due to the Bessel functions with $n$ in the argument. It is somewhat complicated
and we do not reproduce it here since it is not needed: as we already remarked, it is better to keep the Bessel function unexpanded.

One final comment: for $\mu=0$ the seed $\sigma_{\mu=0}$  is a power of $\tau_2$, and the integral and summation over $c$ for the constant term is a power of $\tau_2$, whereas the nonzero modes produce a Bessel function $K_{s-1/2}$. Here for $\mu\neq 0$, the seed is a Bessel function $K_{s}$ and  result of the integration and summation
over $c$ for the nonzero  is a Bessel function $K_{s-1/2}$ both for zero mode and the nonzero modes. So for the massive Eisenstein series, the distinction between zero modes and nonzero modes is less pronounced than for $\mu=0$.

\section{Theta lifts and the Rankin-Selberg-Zagier transform}
\label{ThetaRSZ}
The Rankin-Selberg method is a general approach for constructing new automorphic L-functions by integrating the product of two modular forms over a group. The Rankin-Selberg-Zagier (RSZ) transform makes use of this idea as a way of regularizing a theta-type integral. The basic idea is to deform an integral by inserting a non-holomorphic Eisenstein series in the integrand, thus creating a family of integrands parametrized by $s\in \mathbb{C}$. The original integral is obtained by analytically continuing the result to $s=0$ or $s=1$. We start this section by giving a brief summary of the RSZ-transform, after which we show how it can be applied to calculate theta lifts. In subsequent sections we will then generalize these constructions to the massive theta lifts. 
\subsection{The RSZ-transform}

\begin{figure}[t]
\begin{center}
\includegraphics[width=0.8\textwidth]{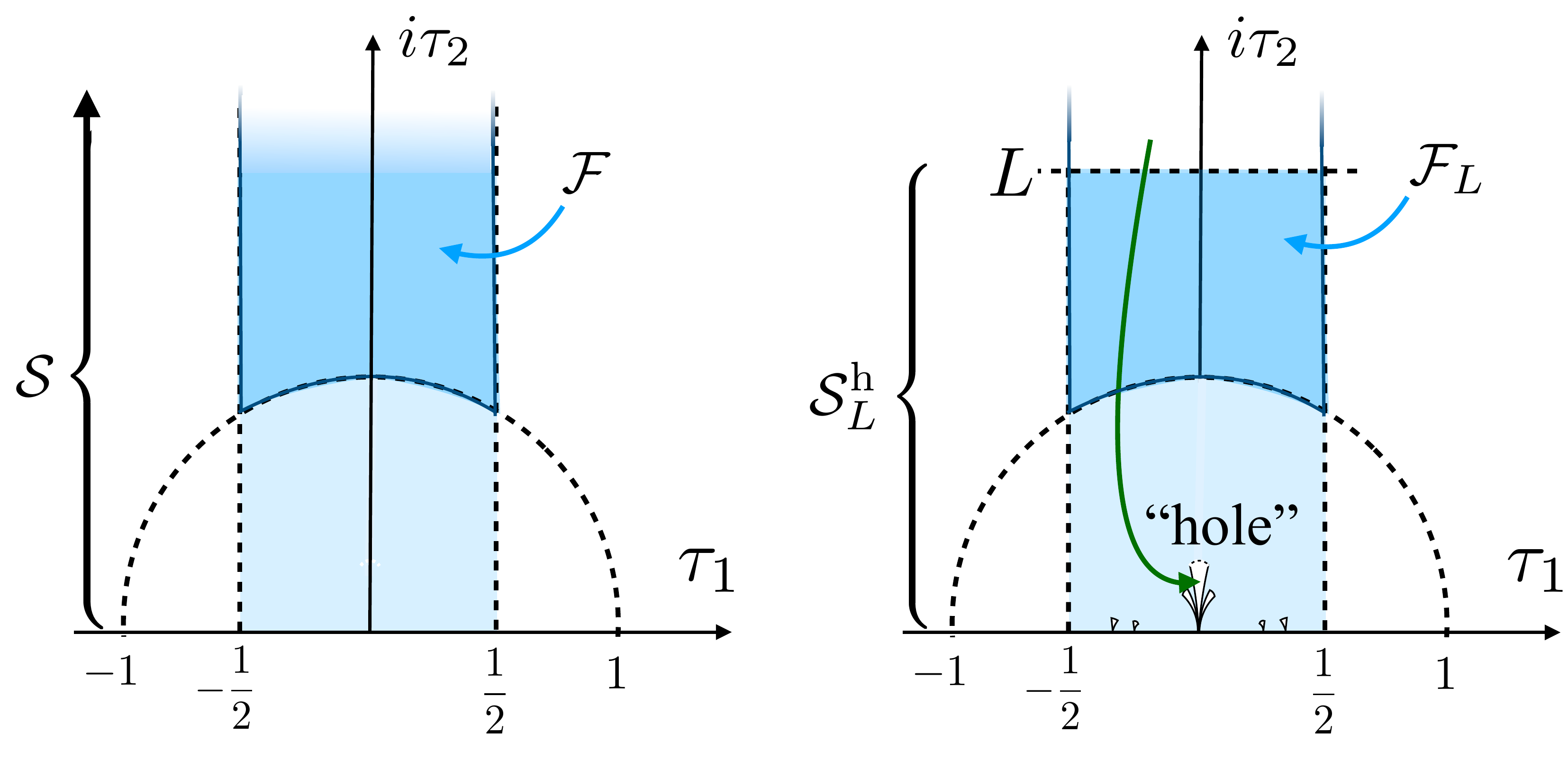}
\caption{Left: full strip  ${\mathcal S} : -1/2<\tau_1<1/2$. Right: cutoff fundamental domain ${\mathcal F}_L$.
The region $\tau_2> L$ above the cutoff maps under $\Gamma$ to ``holes'' in 
the full strip. The strip ${\mathcal S}$ minus the holes is the ``holey'' strip ${\mathcal S}_L^{\rm h}$.  A few representative holes are drawn. }
\label{fig:funda}
\end{center}
\end{figure}

In this section we consider the basic integral of a modular invariant function $F(\tau)$ over the the fundamental domain ${\mathcal F}=SL(2,\mathbb{Z})\backslash \mathbb{H}$:
\begin{equation} \label{starting}
\int_{\mathcal F} F(\tau)  \, {d \tau d\bar{\tau} \over \tau_2^2} \; . 
\end{equation}
For functions that are not of rapid decay, Zagier \cite{Zagier81}  introduced
a cutoff fundamental region ${\mathcal F}_L$ with $\tau_2\leq L$, as in fig.\ \ref{fig:funda}. (See also \cite{Angelantonj:2011br} for a nice physics-oriented review.)
The RSZ transform of an $SL(2,\mathbb{Z})$-invariant function $F(\tau)$ is given by 
introducing an Eisenstein series $E_s(\tau)$ into the integral (for Re $s>1$) and writing it as a Poincar\'e series that
can be realized as a single term by unfolding the integral to the strip:
\begin{equation}
R_{L}(F,s) = \int_{\mathcal F_{L}} F(\tau) E_{s}(\tau) \, {d \tau d\bar{\tau} \over \tau_2^2} 
= \int_{\mathcal F_{L}} F(\tau) \!\!\sum_{\gamma \in \Gamma_{\infty} \backslash \Gamma} ({\rm Im}(\gamma\cdot \tau))^s \, {d \tau d\bar{\tau} \over \tau_2^2} =  \int_{ {\mathcal S}_L^{\rm h} }\!\!F(\tau)({\rm Im}\,  \tau )^s \, {d \tau d\bar{\tau} \over \tau_2^2} \; . 
\end{equation}
For $L\rightarrow \infty$, the domain of integration is unfolded to the full strip ${\mathcal S}$ as in the left panel of figure \ref{fig:funda}. 
However, for finite $L$,
the unfolded domain ${\mathcal S}_L^{\rm h} $ is {\it not} simply the  strip cut off at the top, since  it has ``holes'' due to the cutoff.  We can expand in $1/L$ to isolate divergences, and subtract them to construct a renormalized $R^{\rm ren}_{L}(F,s)$. 
If we now  take $L\rightarrow \infty$, 
the holes are filled in.\footnote{We thank Federico Zerbini for discussions on this point.}

We assume that the modular invariant function $F(\tau)$ has a Fourier expansion in $\tau_1$:
\begin{equation}
F(\tau)=f(\tau_2)+\sum_n g_n(\tau_2)e^{2\pi i n\tau_1} \; . 
 \end{equation}
Since the seed function $({\rm Im}\,  \tau )^s=\tau_2^s$ is independent of $\tau_1$, and
$\int_{-1/2}^{1/2}  e^{2\pi i n\tau_1}d\tau_1=0$ for $n\neq 0$, the $\tau_1$ integral projects out the non-zero mode sum, and we only have the zero mode left:
\begin{equation} \label{result}
R_{\infty}(F,s)=\int_{0}^{_{\infty}}\tau_2^{s-2} f(\tau_2) d\tau_2 \label{transform}
\end{equation}
which  is a Mellin transform of the zero-mode $f$ of $F$. Recall from above that Re $s>1$. 
Note also that the Fourier projection $\int_{-1/2}^{1/2} e^{2\pi i n\tau_1}d\tau_1=0$ for $n\neq 0$ applies for the {\it full} strip ${\mathcal S}$, not for the ``holey'' strip
${\mathcal S}_L^{\rm h}$.

If we are only interested in the original integral eq.\  \eqref{starting} over the fundamental domain, the clever Rankin-Selberg trick is that the Eisenstein series has a pole
at $s=1$ with $\tau_2$-independent residue, so we can extract the value of the integral \eqref{starting} from the residue of eq. \eqref{result}.
However, for some applications the full RSZ transform $R(F,s)$ (i.e.\ including the Eisenstein series $E_s$ in the integrand) is of interest, so we do not view the residue of eq. \eqref{result} as the only interesting property of  the RSZ transform  $R(F,s)$.

In this paper, we mainly integrate functions $F$ that decay exponentially as $\tau_2\rightarrow \infty$, except possibly
for some degenerate piece that can be handled separately. We can then take the $L\rightarrow \infty$ limit
with less effort than for more general falloff conditions. Nevertheless, in a few places we will find it useful to compare to Zagier's approach.

\subsection{Theta Lift}
By the motivations discussed in the introduction, 
we want to Rankin-Selberg transform the $d=2$ Siegel-Narain theta function:

\begin{equation} \label{SNt}
\Theta_{2,2}(\tau)=\tau_2 \sum_{k\in {\mathbb Z}^{2*}\!,n\in {\mathbb Z}^{2}}
\exp\Big({-{\pi  \tau_2}(\underbrace{(k+Bn)G^{-1}(k+Bn)+nGn)}_{\mathcal M}+2\pi i \tau_1 \langle k,n\rangle}\Big) \; . 
\end{equation}
with  real real 2$\times$2 matrices  $G$ (symmetric) and $B$  (antisymmetric). 
The zero mode $f(\tau_2)$ of $F=\Theta_{2,2}$ consists of the terms $(k^1,k^2,n_1,n_2)$ for which  $\langle k,n\rangle=k^1 n_1+k^2 n_2=0$.
The Poincar\'e series seed function for $E_s$ with our normalization is simply $\tau_2^s$, so we have: 
\begin{equation} \label{RSZmassiveT}
R_{L}(\Theta,s) = \int_{\mathcal F_{L}}\Theta_{2,2}(\tau){E}_{s}(\tau) \, {d \tau d\bar{\tau} \over \tau_2^2}
= \int_{{\mathcal S}_L^{\rm h} } \Theta_{2,2}(\tau) \tau_2^s {d \tau_1d\tau_2 \over \tau_2^2}.
\end{equation}
If Re $s>1$, we can send $L\rightarrow \infty$ and perform the $\tau_1$ integral:
\begin{eqnarray} \label{Rmassless}
R_{\infty}(\Theta,s)= \!\!\sum_{{{k\in {\mathbb Z}^{2*}\!,n\in {\mathbb Z}^{2}} \atop \langle k,n\rangle=0} }  \int_0^{\infty} 
\! \tau_2^{s-1} e^{-{\pi  \tau_2}{\mathcal M}}  d\tau_2
 &=& { \Gamma(s) \over \pi^s} \!\!\!\sum_{{{k\in {\mathbb Z}^{2*}\!,n\in {\mathbb Z}^{2}} \atop \langle k,n\rangle=0} }{1 \over \mathcal M^s}
=: E_{V,s} \; ,
\end{eqnarray}
with $E_{V,s}$ the {\it lattice Eisenstein series}. Here ${\mathcal M}$ implicitly depends on complex moduli $T$ and $U$ that come from the real matrices $G$ and $B$: we will make the moduli dependence of $ E_{V,s}$ (and some of its generalizations) explicit in eq.\ \eqref{Evec} below. Here, we only need the property \eqref{EVfactor} below:
for Re $s>1$, the lattice  Eisenstein series $E_{V,s}$ factors into one factor depending on the modulus $T$ and one factor depending on the modulus $U$:
\begin{equation} \label{prodrepeat}
E_{V,s}(T,U) =  \pi^{-s} \Gamma(s)  \zeta(2s)E_s(T)E_s(U) \; , 
\end{equation}
which representation-theoretically comes from $O(2,2) \sim SL(2)\times SL(2)\ltimes {\bf Z}_2$.  
We can take the $s\rightarrow 1$ limit by Kronecker's first limit formula: each $E_s$  factor has a single pole at $s=1$, which in our normalization is eq.\ (10.10) in the book \cite{Fleig:2015vky}:
\begin{equation} \label{K1}
E_s(\tau) = {3 \over \pi(s-1)}-{6 \over \pi }(\log(\sqrt{\tau_2}|\eta(\tau)|^2 + \mbox{constant})  +{\mathcal O}(s-1)
\end{equation}
so by combining eqs.\ \eqref{prodrepeat} and \eqref{K1}, the single pole of the Rankin-Selberg transform at $s=1$ arises as a cross term:
\begin{equation}
E_{V,s}= {3 \over 2\pi(s-1)^2}-{3 \over \pi(s-1)}(\log( \sqrt{T_2U_2}|\eta(T)\eta(U)|^2)+\mbox{constant})
+{\mathcal O}((s-1)^0)
\end{equation}
and by the Rankin-Selberg trick mentioned above,
the  simple pole  gives the moduli dependence of the classic Dixon-Kaplunovsky-Louis threshold correction $\Delta_{\rm DKL}$ \cite{Dixon:1990pc}:
\begin{equation}
\Delta_{\rm DKL}(T,U) ={\pi \over 3}\,  \mathop{\mathrm{Res}}_{s=1} \, R_{\infty}(\Theta,s) = -{1 \over 2}\log( T_2U_2|\eta(T)\eta(U)|^4)+\mbox{constant} \; ,
\end{equation}
that only differs from the original DKL result by  overall normalization and an additive renormalization-scheme dependent constant.

To summarize, since we allowed for generic $s$, i.e.\ allow for poles in the transform, there was no need for cutoff $L$ in the RSZ transform of the Siegel-Narain theta function. This is analogous to dimensional regularization.

Alternatively, we could have fixed $s=1$ and continued with the cutoff $L$, with the result that $\Gamma(s)$ would have been replaced by $\Gamma(s)-\Gamma(s,\pi L {\mathcal M})$,
where $\Gamma(s,x)$ is the (lower) incomplete gamma function and $\Gamma(1,x)=e^{-x}$.
But then the application of the factorization \eqref{prodrepeat} is less direct.

\subsection{Selberg-Poincar\'e series}

 The Siegel-Narain $\Theta$ is a sum of  exponentially suppressed terms
and converges quickly.
In the previous section we encountered its theta lift: the lattice Eisenstein series $E_{\rm V,s}$.
For later use, we will introduce an integer superscript $\ell$ and refer to the previous
case as $\ell=0$:
\be  \label{firstint}
 \int  \Theta(T,U,\tau) E_s^{(\ell)}(\tau){d^2 \tau \over \tau_2^2} = E_{{\rm V},s}^{(\ell)}(T,U) 
\ee
where 
the Selberg-Poincar\'e series \cite{Selberg} is a Poincar\'e series
with seed $\tau_2^s q^{-\ell}$
\be  \label{SelbergPoincare}
E_s^{(\ell)}(\tau) = {1 \over 2\zeta(s)}\sum_{c,d}{\tau_2^s \over |c\tau+d|^{2s}}(\widetilde{q})^{-\ell}
\ee
with $\widetilde{q} = e^{2\pi i \, \widetilde{\tau}} = e^{2\pi i (a\tau+b)/(c\tau+d)}$, and $a,b$ are determined from $ad-bc=1$. 
Clearly for $\ell=0$, $E_s^{(0)}(\tau)=E_s(\tau)$ reduces to the ordinary nonholomorphic Eisenstein series. For $\ell=1$, the Selberg-Poincar\'e series
$E_s^{(1)}(\tau)$ provides a non-holomorphic regularization of Klein's modular invariant $j$ function as a Poincar\'e series,
that becomes holomorphic as $s \rightarrow 0$. 
For $\ell>1$, $E_s^{(\ell)}(\tau)$
can be obtained from Hecke operators acting on $j$ \cite{Angelantonj:2011br}.\footnote{For example for $s=0$, we have $E_0^{(\ell)} = T_{\ell}j + 12 \sigma_1(\ell)$ (see \cite{Angelantonj:2011br} for more details).}
The result of unfolding the Siegel-Narain $\Theta$ against the the Selberg-Poincar\'e series $E_s^{(\ell)}$  in
 \eqref{firstint} is the quadruple sum constrained lattice Eisenstein series
\be \label{quad} \label{Evec}
E_{{\rm V},s}^{(\ell)}(T,U) = \sum_{k\in {\mathbb Z}^{2*}\!,n\in {\mathbb Z}^{2}} { \delta(|p_{\rm L}|^2 - |p_{\rm R}|^2-4\ell) \over (|p_{\rm L}|^2 + |p_{\rm R}|^2-4\ell)^s } \; , \quad {\rm Re}\, s>2
\ee
 where  the momenta $p_{\rm L}$, $p_{\rm R}$ give the dependence on the moduli $T,U$ as
\be  \label{pLpR}
p_{\rm R}={n_2-U n_1 + T k^1 + TU k^2 \over \sqrt{T_2U_2}} \; , \quad p_{\rm L}={n_2-U n_1 + \overline{T} k^1 + \overline{T} U k^2 \over \sqrt{T_2U_2}} \; . 
\ee
As noted,  the quadruple sum in  \eqref{Evec} generically only converges for Re $s>2$. In particular, unlike for the ordinary nonholomorphic Eisenstein series, if we are interested in $s=1$,
we cannot set $s=1+\epsilon$ with $\epsilon$ small for convergence.
 In the following section we view \eqref{firstint} as an integral representation
that analytically continues the the Selberg-Poincar\'e series \eqref{quad} to all $s$. 

In the remainder of this section, we review how the constraint $|p_{\rm L}|^2 - |p_{\rm R}|^2-4\ell=0$ in \eqref{Evec} arises,
how to perform the sum, and give the moduli dependence in more detail.
In complex coordinates $T,U$,  the Siegel-Narain theta function \eqref{SNt} becomes
 \be
 \Theta(T,U,\tau) = \tau_2\sum_{k\in {\mathbb Z}^{2*}\!,n\in {\mathbb Z}^{2}} q^{{1 \over 4}|p_{\rm L}|^2 }\bar{q}^{{1 \over 4}|p_{\rm R}|^2 }
  =  \tau_2\sum_{k\in {\mathbb Z}^{2*}\!,n\in {\mathbb Z}^{2}} e^{{\pi i \tau_1 \over 2}(|p_{\rm L}|^2-|p_{\rm R}|^2)}e^{-{\pi \tau_2 \over 2}(|p_{\rm L}|^2+|p_{\rm R}|^2)}
 \ee
and we note that the coefficient of $\pi i \tau_1/2$ in the exponent is
\be
|p_{\rm L}|^2-|p_{\rm R}|^2 = -4\langle k, n\rangle=-4(k^1 n_1+k^2 n_2) \; . 
\ee
The $\tau_1$ integral in \eqref{firstint} becomes
(essentially giving the $\ell$th Fourier coefficient of $\tau_1$ of the Siegel-Narain theta function $ \Theta$) 
\be
\int_{-1/2}^{1/2} d\tau_1\,  e^{2\pi i \ell \tau_1} \Theta(\tau) =\int_{-1/2}^{1/2} d\tau_1\,   \tau_2e^{2\pi i \ell \tau_1} e^{2\pi i \tau_1 \langle k, n\rangle}
\!\!\!\!\!\!\sum_{k\in {\mathbb Z}^{2*}\!,n\in {\mathbb Z}^{2}}\!\!\!\!\tau_2e^{-{\pi \over 2} \tau_2 (|p_{\rm L}|^2 + |p_{\rm R}|^2)} = 
\!\!\sum_{\langle k, n\rangle= -\ell}\!\!\!\!\tau_2e^{-{\pi \over 2} \tau_2 (|p_{\rm L}|^2 + |p_{\rm R}|^2)}
\ee
so we have a quadruple sum over $(k^1,k^2,n_1,n_2)$ with a single constraint $\langle k, n\rangle=k^1 n_1+k^2 n_2= -\ell$.
We can perform the quadruple sum while enforcing this condition, or we can solve the constraint as discussed below.

Finally including also the Selberg-Poincar\'e series in \eqref{firstint} we have
\be \label{explicit}
 \int  \Theta(T,U,\tau) E_s^{(\ell)}(\tau){d^2 \tau \over \tau_2^2}
 =\int_0^{\infty}\tau_2^{s-1} e^{2\pi \ell \tau_2} 
 \!\!\sum_{\langle k, n\rangle= -\ell}\!\!\!\!e^{-{\pi \over 2} \tau_2 (|p_{\rm L}|^2 + |p_{\rm R}|^2)}
 d\tau_2 .
\ee
As already emphasized, we cannot move the sum out of the integral for $s\leq 2$, which for many purposes is the region of interest.
We discuss analytic continuation in  appendix \ref{analytic4}.

\subsubsection{Degenerate constraint $\langle k,n\rangle = 0$}
We need to solve the summation constraint $\langle k,n\rangle = k^1 n_1+k^2 n_2=0$. There are two cases:

{\bf Case 1}. If $k^1=k^2=0$, $n_1,n_2$ are unconstrained.
We find
\be
|p_{\rm L}|^2+|p_{\rm R}|^2 = {2 \over T_2 U_2}| n_1 U -n_2|^2 \; . 
\ee
This gives a contribution to \eqref{explicit} that is
\begin{eqnarray} 
\int_0^{\infty}\tau_2^{s-1} 
 \!\!\sum_{n_1,n_2}\!\!e^{-{\pi \over 2} \tau_2 (|p_{\rm L}|^2 + |p_{\rm R}|^2)}
 d\tau_2=
\int_0^{\infty}\tau_2^{s-1}  
 \!\!\sum_{n_1,n_2}\!\!e^{-  {\pi\tau_2 \over T_2 U_2}| n_1 U -n_2|^2}\nonumber \\
  \stackrel{*}{=} \pi^{-s}\Gamma(s) T_2^s\sum_{n_1,n_2}\!\!  {U_2^ s \over | n_1 U -n_2|^{2s}}=\pi^{-s}\Gamma(s) \zeta(2s) T_2^s
  E_s(U)  \label{eqET1}
\end{eqnarray}
where $\stackrel{*}{=}$ assumes ${\rm Re}\, s>1$. The $\zeta(2s)$ comes in because
our $E_s(U)$ is defined for coprime summation integers.
Now, \eqref{eqET1} is only a single term in the $(k^1,k^2)$ double sum, so  naturally it has the convergence
 properties of a double sum in $(n_1,n_2)$, as opposed to a quadruple sum over $(k^1,k^2,n_1,n_2)$. In  appendix \ref{sec:analytic} we review the textbook 
 calculation of giving an analytic continuation for double sums, instead of the step labelled ``$\stackrel{*}{=}$''.

{\bf Case 2}. 
View $n_1$ and $n_2$ as fixed. If they are not coprime, we factor out the biggest common factor
 $(n_1,n_2) = c(n_1^*,n_2^*)$, where $c$ can be positive or negative. Now $k^1 n_1+k^2 n_2=0$ becomes $k^1 n_1^*+k^2 n_2^*=0$ (notice this is not true for the nondegenerate constraint $\langle k,n\rangle =-\ell\neq 0$), and we have
\be
k^1 = -{k^2 n_2^* \over n_1^*} \; . 
\ee
Since $(n_1^*,n_2^*)$ have no common factors, for this to be integer we must have
$k^2 / n_1^*=d$ integer for $d\geq 1$. (Including also $k^2 = -d n_1^*$ would overcount,
and $d=0$ is covered by Case 1.)

To summarize, $(k^1,k^2) =  d(-n_2^*,n_1^*)$ and $(n_1,n_2) = c(n_1^*,n_2^*)$. We find
that the exponent factorizes:
\be  \label{pLRfactor}
|p_{\rm L}|^2+|p_{\rm R}|^2 = {2 \over T_2 U_2} | n_2^* U -n_1^*|^2 | d T -c|^2.
\ee
This gives a contribution to \eqref{explicit} that is
\be  \label{eqET2}
\int_0^{\infty}\tau_2^{s-1}
 \!\!\sum_{d\geq 1,c}\sum_{{\rm gcd}(n_1^*,n_2^*)=1}\!\!e^{-  {\pi\tau_2 \over T_2 U_2}| n_2^* U -n_1^*|^2| d T -c|^2}
 d\tau_2
\stackrel{*}{=}  \pi^{-s}\Gamma(s)\sum_{d\geq 1,c}{T_2^s \over | d T -c|^{2s}} \!\!\!\!\!\sum_{{\rm gcd}(n_1^*,n_2^*)=1}\!\!  {U_2^ s \over | n_2^* U -n_1^*|^{2s}}
 \ee
 where again $\stackrel{*}{=}$ only holds for Re $s>1$.
 Adding \eqref{eqET1} and \eqref{eqET2} fills in the missing $d=0$ term to make $E_s(T)$ and the total is
\be   \label{EVfactor}
 \int {d^2 \tau \over \tau_2^2} \Theta(T,U,\tau) E_s^{(0)}(\tau)
 = E_{{\rm V},s}^{(0)}(T,U)  =  \pi^{-s}\Gamma(s) \zeta(2s) E_s(T)E_s(U)  = E^{\star}_s(T)E_s(U)\; ,
 \ee
where the factor $\zeta(2s)$ came from  \eqref{eqET1}, and similarly for \eqref{eqET2}.

We can also consider the  completed transform, that includes a factor of $\xi(2s)=\pi^{-s}\Gamma(s)\zeta(2s)$ in front,
that makes both Eisenstein series completed:
\be   \label{EVfactor2}
 \xi(2s)\int  \Theta(T,U,\tau) E_s^{(0)}(\tau){d^2 \tau \over \tau_2^2}
 = \xi(2s)E_{{\rm V},s}^{(0)}(T,U)   = E^{\star}_s(T)E^{\star}_s(U)\; . 
 \ee
The additional factor $\xi(2s)$ supplies a double pole at $s=0$. The expression is now  symmetric  under application of the functional relation: $E^{\star}_{1-s}=E^{\star}_s$, which implies for the lattice Eisenstein series that
\be \label{refl1}
\xi(2(1-s))E_{{\rm V},1-s}^{(0)}(T,U)  = \xi(2s)E_{{\rm V},s}^{(0)}(T,U).
\ee
Using the functional relation of the completed Riemann zeta function $\xi(s)=\xi(1-s)$ we can also write this in the more familiar form from the $SL(2,\mathbb{R})$ non-holomorphic Eisenstein series:
\be
E_{{\rm V},s}^{(0)}(T,U)=\frac{\xi(2s-1)}{\xi(2s)} E_{{\rm V},1-s}^{(0)}(T,U).
\ee

\subsubsection{Non-degenerate case $\langle k,n\rangle = -\ell$ for $\ell>0$}

In the non-degenerate case $\langle k,n\rangle=-\ell$  we solve the constraint as follows.

Given coprime $(n_1^*,n_2^*)$, the solution of $\langle k,n\rangle =-\ell$
is $n_1=n_1^* + Mk^2$, $n_2=n_2^* - Mk^1$, since
\be
k^1 n_1 + k^2 n_2 = k^1(n_1^* + Mk^2) + k^2(n_2^* - Mk^1) = k^1 n_1^* + k^2 n_2^* = -\ell.
\ee
From the point of view of solving, this is quite different from above. We will see in the next section
that the fiducial (starred) solution appears in the inconsequential upper elements of the $SL(2,{\mathbb Z})$ image,
whereas above, they were summed over and of consequence as the 
$SL(2,{\mathbb Z})$ sum producing the Eisenstein series of $U$.

(The opposite constraint $\langle k,n\rangle=\ell$ formally has the same solution $k_1=k_1^* + Mn^2$, $k_2=k_2^* - Mn^1$ as above,
but the numbers that come out are different.)

\begin{figure}
\begin{center}
\includegraphics[width=0.35\textwidth]{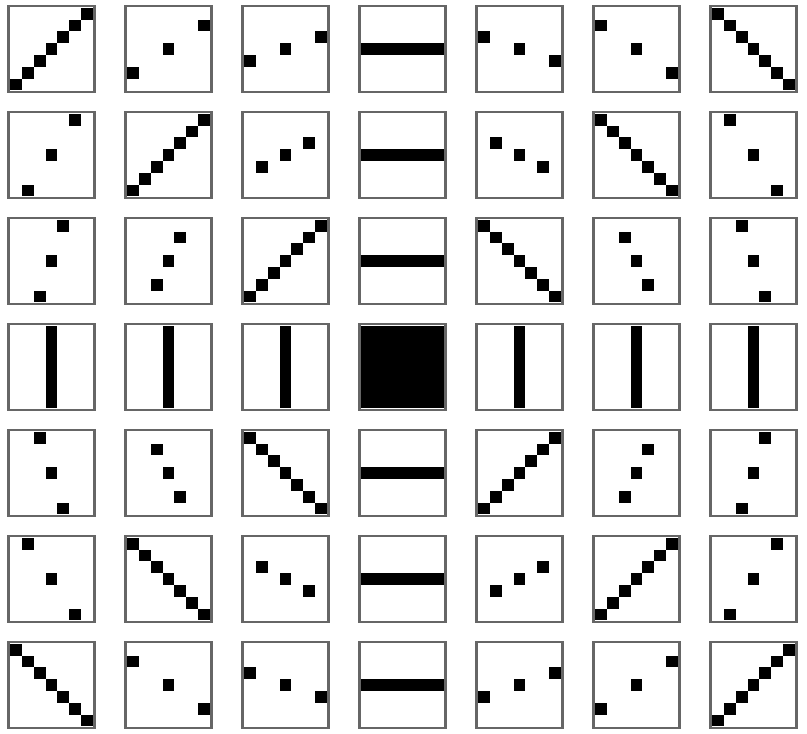}\hspace{7mm}
\includegraphics[width=0.35\textwidth]{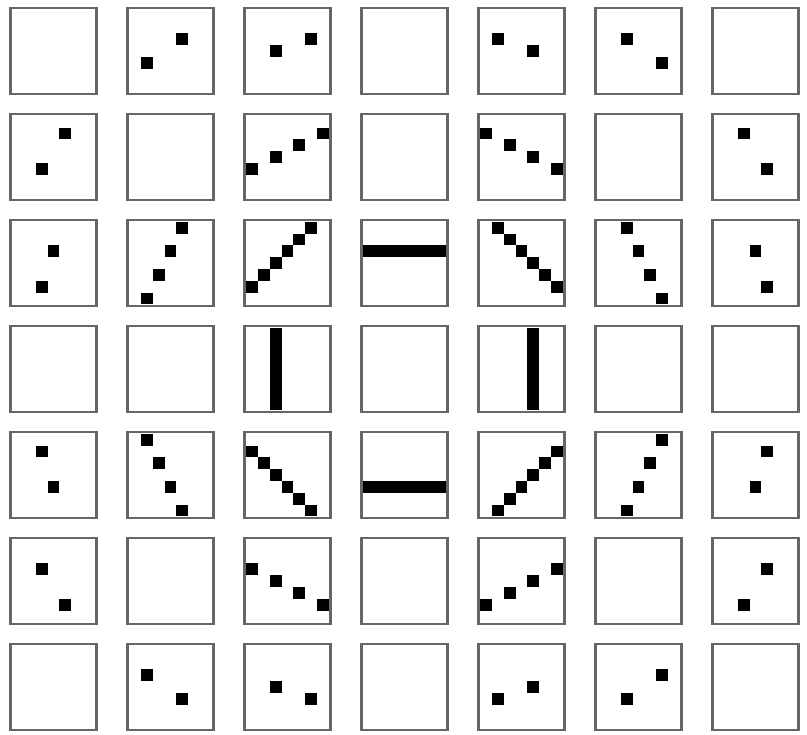}
\end{center}
\caption{ Each element in the 7 $\times$ 7 $(n_1,n_2)$ matrix is a 7 $\times$ 7 $(k^1,k^2)$ matrix. 
The black squares are where the constraint  is satisfied: 
$\langle k,n\rangle=0$ in the left panel and $\langle k,n\rangle=-1$ in the right panel.}
 \label{fig:nondeg2}
\end{figure}

In figure \ref{fig:nondeg2} we show the entries that are included. The constraint $\langle k,n\rangle=-\ell$
is a plane in 4 dimensions. In each truncated little 7 $\times$ 7 block in the figure,
the constraint is satisfied along a straight line. The angle of this plane is different for different blocks, hence the appearance
of ``circles''. We will have occasional use of truncations of our infinite sums, and these graphical representation can provide
some visual checks that truncated summations include a sufficient number of terms in the right range.

\subsection{Theta lift of the $j$-function}
\label{jfunction}
Now we want to RSZ-transform $j(\tau)\Theta_{2,2}(\tau)$:
\begin{equation}
j(\tau)\Theta_{2,2}(\tau)=\tau_2\sum_{\ell=-1}^{\infty}c(\ell)e^{-2\pi \ell \tau_2} e^{2\pi i \ell \tau_1}
\!\!\!\!\!\! \sum_{{k\in {\mathbb Z}^{2*}\!,n\in {\mathbb Z}^{2}}}\!\!\!\!\!\!
\exp\Big({-{\pi  \tau_2}(\underbrace{(k+Bn)G^{-1}(k+Bn)+nGn)}_{\mathcal M}+2\pi i \tau_1 \langle k,n\rangle}\Big) \; ,
\end{equation}
so the zero mode is not given by $\langle k,n\rangle=0$ as it was without the $j$, but now by $\langle k,n\rangle=-\ell$. 
Note that the $\ell=0$ term is the same as before up to an overall constant, so we can set $c(0)=0$ for now,
and $j$ with $c(0)=0$ is denoted $J$. (We return to $c(0)$ below.)
We find
\be \label{RSZT}
R_{L}(J \Theta,s) = \int_{\mathcal F_{L}}J(\tau)\Theta_{2,2}(\tau){E}_{s}(\tau) \, {d \tau d\bar{\tau} \over \tau_2^2}
= \int_{\mathcal S_L^{\rm h}}J(\tau) \Theta_{2,2}(\tau) \tau_2^s {d \tau_1d\tau_2 \over \tau_2^2}  \; . 
\ee
Taking $L\rightarrow \infty$, and provided we can use the $q$ expansion of $J$:
\be
  \!\!\int_0^{ \infty} \sum_{\ell=-1}^{\infty}c(\ell)e^{-2\pi \ell \tau_2} \tau_2 \! \sum_{{{k\in {\mathbb Z}^{2*}\!,n\in {\mathbb Z}^{2}} \atop \langle k,n\rangle =-\ell} }
e^{-{\pi  \tau_2}{\mathcal M}}    \tau_2^{s-2} d\tau_2
=\sum_{\ell=-1}^{\infty} c(\ell) \sum_{{{k\in {\mathbb Z}^{2*}\!,n\in {\mathbb Z}^{2}} \atop \langle k,n\rangle =-\ell} }  \!\int_0^{\infty} 
\! \tau_2^{s-1} e^{-{\pi  \tau_2}({\mathcal M}+2\ell)}  d\tau_2 \label{Jlift}
\ee
The $\ell=-1$ term (the $1/q$ pole in $j$) appears to cause the integral to diverge, but  the minimum value of ${\mathcal M}$ for specific moduli is
2, so for generic moduli ${\mathcal M}>2$, eq.\   \eqref{Jlift} still falls off rapidly at $\tau_2 \rightarrow \infty$ as it did without the $J$ function. However, the $\tau_2\rightarrow 0$ side is unsuppressed.

There is the {\it singular theta lift}
to take care of this new divergence \cite{Harvey:1995fq,BorcherdsSingular}. From the physics point of view,
the main feature is the Harvey-Moore threshold correction
\begin{equation} \label{RSZjmassiveT}
\Delta_{\rm HM}(T,U) = {\pi \over 3}
\mathop{\mathrm{Res}}_{s=0}  R_{\infty}(J \Theta,s) = -2\log |J(T)-J(U)|^2 \; . 
\end{equation}
This is not regularized for $T=U$.

\section{Massive theta lift}
After this brief review of standard theta lifts, we are ready to introduce the massive theta lift. We  consider the theta integral with the insertion of a massive Eisenstein series, in the spirit of the RSZ transform. We start by discussing the general structure of such integrals, after which we explicitly calculate the lift of the constant function, and in a later section a $j$-lift. 
\label{massivethetalift}

\subsection{Massive RSZ in general}
We replace the Eisenstein series $E_s$ in the standard RSZ transform \eqref{RSZmassiveT} with the massive Eisenstein series ${\mathcal E}_{s,\mu}$ from \cite{Berg:2019jhh}. Like for massless, massive RSZ unfolds the integral to the strip with holes ${\mathcal S}_L^{\rm h}$:
\be \label{RSZmassive}
R_{L,\mu}(F,s) = \int_{\mathcal F_{L}} F(\tau) {\mathcal E}_{s,\mu}(\tau,\bar{\tau}) \, {d \tau d\bar{\tau} \over \tau_2^2} = {\pi^s \over \Gamma(s)} \int_{\mathcal S_L^{\rm h}} F(\tau) 2(\mu\tau_2)^{s/2}K_s\left(2\pi \sqrt{\mu/\tau_2}\right)  {d \tau_1d\tau_2 \over \tau_2^2} 
\ee
and if we take the $L\rightarrow \infty$ limit, we can trivially perform the $\tau_1$ integral, like above:
\be
R_{\infty,\mu}(F,s) = {2 \pi^s  \mu^{s/2} \over \Gamma(s)}  \!\!\int_0^{\infty}    \tau_2^{s/2-2}K_s\left(2\pi \sqrt{\mu/\tau_2}\right) f(\tau_2) d\tau_2 \nonumber
\ee
This is somewhat like a  Bessel transform, rather than a Mellin transform, but note that $\tau_2$ is in the denominator in the argument of the Bessel function.

A few words on asymptotics. For $\tau_2\rightarrow 0$, because $\tau_2$ appears in the denominator in the argument of the Bessel function, the  $\mu$ deformation improves the behavior of the seed: it goes as $e^{-2\pi\sqrt{\mu/\tau_2}}\tau_2^{s/2-7/2}$
rather than diverging.  For $\tau_2 \rightarrow \infty$, on the other hand, and for fixed $\mu$, the deformation does not help,
and the seed function diverges as $\tau_2^s$,
like it does for the usual massless case $\mu=0$. Subleading terms now depend on $\mu$, and depending on $s$ they are either divergent or not. 
 
Note also that although the seed is finite at $\tau_2\rightarrow0$, it is nonanalytic in $\tau_2$ due to the $e^{-2\pi\sqrt{\mu/\tau_2}}$ factor. A theme in this paper is that nonanalyticity can make some calculations appear  unfamiliar, but if it helps regulate some divergences,
it could be worth it.

\subsection{The massive theta lift}
With nonzero mass parameter $\mu$, we will not need to dimensionally regularize
to complex $d$, or zeta-regularize to complex $s$. This can be an advantage in some contexts, for example since dimensional regularization generically does not respect supersymmetry.

The Siegel-Narain theta function is modular invariant. (This is more obvious in the Poisson-resummed form, \cite{Fleig:2015vky} eq.\ (13.135)). 
Again, the zero mode $f(\tau_2)$ of $F=\Theta_{2,2}$ consists of the terms when $\langle k,n \rangle=0$:
\begin{equation}
f(\tau_2)=\tau_2 \sum_{{k\in {\mathbb Z}^{2*}\!,n\in {\mathbb Z}^{2} \atop \langle k,n \rangle =0} }
e^{-{\pi  \tau_2}{\mathcal M}}.
\end{equation}
One could consider whether $\Theta$ should itself be deformed by $\mu$. The DKL correction would then have massive Dedekind $\eta$ functions of $T$ and $U$. 
However, at least in the examples of plane gravitation wave backgrounds \cite{Bergman:2002hv}, the massive worldsheet fields $X^{m}$ appear in the external directions polarized along the wave (usually
viewed to be noncompact),
whereas the $T$, $U$ moduli parametrize an internal torus $X^i$ (which is compact). In this product geometry, the Kaluza-Klein and winding modes 
along the internal torus $X^i$  are unaffected by the mass deformation in the noncompact directions. One could consider a non-product geometry, but that is beyond the scope of this work. 

On the other
hand, as we will see, 
even though we do not begin with a massive ${\Theta}$, we end up with a massive
function of the spacetime moduli. 

\begin{figure}[h]
\begin{center}
\includegraphics[width=0.45\textwidth]{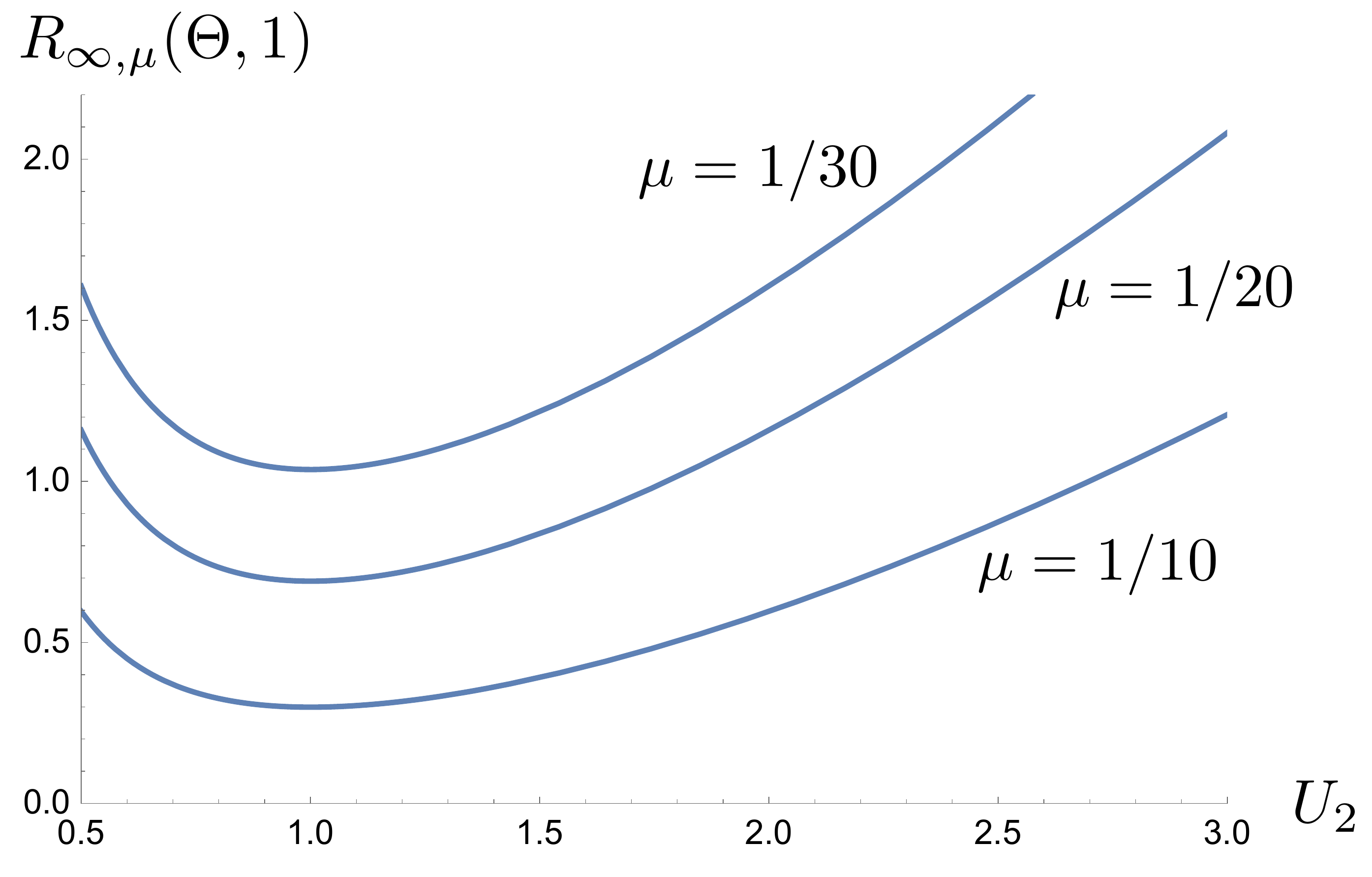}
\caption{Plot of \eqref{expl} for $T=i$, $U=iU_2$. The sum was truncated to $|k^1|,|k^2|,|n_1|,|n_2| \leq 5$, which with
$\langle k,n \rangle =0$ leaves 832 terms, of which 99 distinct.
More terms do not change this plot.}
 \label{Gplot}
 \end{center}
\end{figure}

The massive Rankin-Selberg transform of $F=\Theta_{2,2}$  is
\be \label{RSZmassiveT}
R_{L,\mu}(\Theta,s) = \int_{\mathcal F_{L}}\Theta_{2,2}(\tau){\mathcal E}_{s,\mu}(\tau) \, {d \tau d\bar{\tau} \over \tau_2^2}
= {\pi^s \over \Gamma(s)} \int_{\mathcal S_L^{\rm h}} \Theta_{2,2}(\tau) 2(\mu\tau_2)^{s/2}K_s\left(2\pi \sqrt{\mu/\tau_2}\right)  {d \tau_1d\tau_2 \over \tau_2^2} 
\ee
If we send $L\rightarrow \infty$, we can integrate over $\tau_1$ and find
\begin{eqnarray}
R_{\infty,\mu}(\Theta,s)&=&{2\pi^s  \mu^{s/2}  \over \Gamma(s)} \!\!\int_0^{\infty}  \tau_2 \sum_{k\in {\mathbb Z}^{2*}\!,n\in {\mathbb Z}^{2} \atop \langle k,n \rangle =0}
e^{-{\pi  \tau_2}{\mathcal M}}    \tau_2^{s/2-2}K_s\left(2\pi \sqrt{\mu/\tau_2}\right)d\tau_2 \nonumber \\
&=& \sum_{k\in {\mathbb Z}^{2*}\!,n\in {\mathbb Z}^{2} \atop \langle k,n \rangle =0} {2\pi^s  \mu^{s/2} \over \Gamma(s)}  \!\!\int_0^{\infty} 
 \tau_2^{s/2-1} e^{-{\pi  \tau_2}{\mathcal M}}   K_s\left(2\pi \sqrt{\mu/\tau_2}\right)d\tau_2 \label{Kintegral} \nonumber\\
 &=&  \sum_{k\in {\mathbb Z}^{2*}\!,n\in {\mathbb Z}^{2} \atop \langle k,n \rangle =0}{2\pi^s  \mu^{s/2} \over \Gamma(s)} \cdot  {\pi^s \over 2}\mu^{s/2}
 G^{3,0}_{0,3}(0,0,-s,\pi^3 \mu  {\mathcal M})\nonumber\\
 &=& {\pi^{2s} \over \Gamma(s)}  \sum_{k\in {\mathbb Z}^{2*}\!,n\in {\mathbb Z}^{2} \atop \langle k,n \rangle =0}  \mu^s G^{3,0}_{0,3}(0,0,-s,\pi^3 \mu  {\mathcal M}) \; .  \label{expl}
\end{eqnarray}
Here $G$ is the Meijer G function, a type of Mellin-Barnes representation:
\begin{equation} \label{intG}
G^{3,0}_{0,3}(b_1,b_2,b_3,z) = {1 \over 2\pi i}\int_C \Gamma(b_1+t)\Gamma(b_2+t)\Gamma(b_3+t) z^{-t}dt
\end{equation}
that we review in appendix \ref{Meijer}. As described there in greater detail, the contour $C$ avoids the poles. In particular, for the form of the arguments in \eqref{expl} we have
\begin{equation} \label{intG2}
G^{3,0}_{0,3}(0,0,-s,z) = {1 \over 2\pi i}\int_C \Gamma(t)^2\Gamma(t-s) z^{-t} dt \;. 
\end{equation}
Here the contour $C$ passes to the right of poles, for example we can pick $t=3/2+ix$ and $x\in (-\infty,\infty)$.
We plot \eqref{expl} for $s=1$ in fig.\ \ref{Gplot}.

The Meijer G function  can also be written as a  sum of hypergeometric functions. We review in appendix \ref{Meijer} why a sum of hypergeometric functions is a less useful representation for our purposes. Briefly, each term in the sum may require separate regularization at special parameter values, whereas the Meijer G function represents a ``coherent'' regularization among those terms, in a sense. However, probably everything in this paper could be reexpressed in terms of somewhat lengthier expressions with hypergeometric functions.

\begin{figure}
\begin{center}
\includegraphics[width=0.35\textwidth]{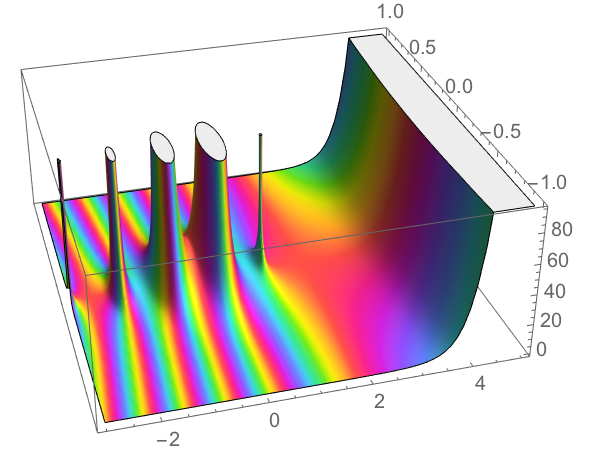}
\caption{The coefficient $\Gamma(t)^2\Gamma(t-s)$ of $z^t$ in the integrand  eq.\ \eqref{intG2}, for
$s=1$: a single pole at $t=1$ and triple poles at $t=-n$ for $n=0,1,2,\ldots$.}
\end{center}
 \label{nondeg2}
\end{figure}

Although we will not only be interested in small $z$, for illustration let us consider  small $z$ so we
can truncate the sum over residues. Again for illustration, set $s=1$ directly:
\begin{equation} \label{intG2}
G^{3,0}_{0,3}(0,0,-1,z) =\mathop{\mathrm{Res}}_{t=1,0,-1,\ldots}\Gamma(t)^2\Gamma(t-s) z^{-t}
={1 \over z} -{1 \over 2}\log^2 z +(1-3\gamma) \log z + \mbox{constant} +{\mathcal O}(z)
\end{equation}
Compare  \eqref{intG2} and  \eqref{expl}.
The argument $z$ in \eqref{expl} contains the mass parameter $\mu$.
As already remarked, expanding around small $\mu$ ruins some properties of the massive Eisenstein series,
so we will only use this expansion to check the connection to the undeformed case. Still, we wanted to illustrate that the expansion \eqref{intG2}
 is explicit and unambiguous for each term in the sum \eqref{expl}. 
 
Now the more important question what happens at fixed $\mu$ and large summation variables, so large ${\mathcal M}$, and therefore large argument $z$ of the Meijer G function. At $z\rightarrow \infty$, the Meijer $G^{3,0}_{0,3}$ is exponentially suppressed but nonanalytic, as
\begin{equation} \label{exprsupp}
G^{3,0}_{0,3}(0,0,-s,z) = {4 \pi \over \sqrt{3}} z^{-(s+1)/3}e^{-3 z^{1/3} +{ \mathcal O}(x^{-1/3})} \; .
\end{equation}
In the massless case ($\mu=0$), the result for the theta lift integral was only power-law suppressed ${\mathcal M}^{-s}$,
despite the factor $e^{-\pi\tau_2  {\mathcal M}}$ in the integrand.
As is discussed in greater detail in the appendix, the exponential suppression \eqref{exprsupp}
can be understood from a change of variables $\tau_2{\mathcal M} = \hat{\tau}_2$
in the integral \eqref{Kintegral}
\be \label{changeofvar}
\int_0^{\infty} 
 \tau_2^{s/2-1} e^{-{\pi  \tau_2}{\mathcal M}}   K_s\left(2\pi \sqrt{\mu/\tau_2}\right)d\tau_2
 ={\mathcal M}^{-s/2}\int_0^{\infty} 
\hat{\tau}_2^{s/2-1} e^{-{\pi \hat{ \tau}_2}}   K_s\left(2\pi \sqrt{\mu \over \tau_2}{\mathcal M}^{1/2} \right)d\hat{\tau}_2
\ee
together with the fact that the Bessel function is exponentially suppressed for large arguments. Without
the Bessel function to ``catch half'' of the ${\mathcal M}$, the change of variables $\tau_2{\mathcal M} = \hat{\tau}_2$ simply brings ${\mathcal M}^{-s}$ out of the integral, leaving the massless integral with  only power-law suppression
of the quadruple sum over $k,n$. (In the massless case, as an alternative,
we can prevent ${\mathcal M}$ 
slipping out of the integral by splitting the integration region $[0,\infty]$  to $[0,1]$ and $[1,\infty]$, as in appendix \ref{sec:analytic}. While expliclit, this has its own challenges, as we will return to in section \ref{alternative}.)

For each term in the quadruple sum, we could also consider expanding around a smooth point like $\mu=1$. 
The function $G^{3,0}_{0,3}(0,0,-s,wz)$ has a smooth series expansion around $\mu=1$, by the multiplication relation
\be 
G^{3,0}_{0,3}(0,0,-s,wz) = w^{-s} \sum_{n=0}^{\infty}{(w-1)^m \over m!}G^{3,0}_{0,3}(0,0,-s+m,z).  \label{mu1}
\ee
One way to use this is to write the sum \eqref{expl} as an infinite sum over $E_{V,s}$ for different integer $s$. Although we will not explore such representations in detail in this paper, we will return to sums of this form in section \ref{mellindual}.

Finally, we see how the massive theta lift \eqref{expl} reductes to the standard theta lift \eqref{Rmassless} for $\mu=0$ as follows.
The factor $\Gamma(t-s)$ has its first pole at $t=s$ (counting from the right), 
that gives $z^{-s}$. Noting that there is an extra factor $\mu$ in front of the Meijer G function in \eqref{expl}, we see that 
\be
R_{\infty,\mu\rightarrow 0}(\Theta,s) ={\Gamma(s) \over \pi^s} \sum_{k\in {\mathbb Z}^{2*}\!,n\in {\mathbb Z}^{2} \atop \langle k,n \rangle =0}
 {1 \over   {\mathcal M}^s} =E_{{\rm V},s}^{(0)}(T,U) 
 \ee
as in \eqref{Rmassless}. 

\subsection{Mellin-dual massive theta lift}
\label{mellindual}
Here we do something perhaps less obvious. 
Let us Mellin-transform with respect to the mass parameter $\mu$, following section 4.2 of \cite{Berg:2019jhh}.
(Of course,  in the massless case ($\mu =0$) there is no direct analog of this operation.)
The variable that is Mellin-dual to the mass parameter $\mu$ is called $t$. We  obtain
\begin{eqnarray} \label{RSZmassiveTM}
{\mathbf M}(R_{L,\mu}(\Theta,s))(t) &=& \int_0^{\infty} \mu^{t-1} \left(\int_{\mathcal F_{L}}\Theta_{2,2}(\tau){\mathcal E}_{s,\mu}(\tau) \, {d \tau d\bar{\tau} \over \tau_2^2}\right) d\mu\nonumber \\
&=& \int_{\mathcal S_L^{\rm h}} \Theta_{2,2}(\tau) \pi^{-2t}{\Gamma(t)\Gamma(s+t) \over \Gamma(s)} \tau_2^{s+t-2}  {d \tau_1d\tau_2 \over \tau_2^2}\nonumber \\
&=&\pi^{-2t}{\Gamma(t)\Gamma(s+t) \over \Gamma(s)} \sum_{k\in {\mathbb Z}^{2*}\!,n\in {\mathbb Z}^{2} \atop \langle k,n \rangle =0} \int_{\mathcal S_L^{\rm h}} \tau_2 
e^{-{\pi  \tau_2}{\mathcal M}}  \tau_2^{s+t-2}  {d \tau_1d\tau_2 \over \tau_2^2} \; . 
\end{eqnarray}
It will be useful for later to note that since here $s$ is just an external parameter as far as this integral transform is concerned, there is freedom to pick some $s$-dependent combination like $s+t$ as the Mellin-dual variable to $\mu$, instead of just $t$. 

Sticking to the choice $t$ and taking $L\rightarrow \infty$, we have
\begin{eqnarray}
{\mathbf M}(R_{\infty,\mu}(\Theta,s))(t)  &=&\pi^{-s-3t} {\Gamma(t)\Gamma(s+t)^2 \over \Gamma(s)}\!\!\!\sum_{k\in {\mathbb Z}^{2*}\!,n\in {\mathbb Z}^{2} \atop \langle k,n \rangle =0}\!\!\!{\mathcal M}^{-(s+t)}  \label{res0}\nonumber \\
&=& \pi^{-2t} {\Gamma(t)\Gamma(s+t) \over \Gamma(s)} E_{V,s+t}({\mathcal M})\; . \label{res}
\end{eqnarray}
As expected from the general arguments of \cite{Berg:2019jhh}, in the Mellin-dual variable $t$, the result looks formally similar to the ordinary massless $\mu=0$ result, except that the latter is divergent for $L\rightarrow \infty$ (as $L^t$) without explicit subtractions.

To Mellin-transform back $t \rightarrow \mu$, we should pick a contour.
This should in principle reproduce the Meijer G function \eqref{intG2} from the previous subsection, but it is not completely obvious at this point
how it does, since the Gamma functions appear in a slightly different combination: $\Gamma(t)\Gamma(s+t)^2$ in \eqref{res0} versus
$\Gamma(t)\Gamma(t-s)^2$ in  \eqref{intG2}. (Remember the contour integral is with respect to $t$, not $s$.)

Viewed as an integral representation of the $\mu$ dependence, we of course do not expect it to be unique. As already mentioned, we can change variables of integration to combinations like $t+s$ or $t-s$ instead of $t$ as Mellin-dual variable. A discrete freedom of parameterization is captured by the functional relation for $E_{V,s+t}$, 
as  we will discuss below.

Like in the previous section, but unlike in the massless case, it is possible to set $s=1$ at this stage. We note that this prevents viewing the massive theta lift as ``infinitesimal'', as we expect it to diverge if we send $\mu\rightarrow 0$, from being on top of the double pole of $ E_{V,s}$.  However, when viewed as a finite deformation this is no problem. 

Now, instead of being on top of the $s=1$ pole, we are faced with $E_{V,1+t}$, that has a double pole at the origin $t=0$ of the complex $t$ plane. The Mellin contour is {\it defined}  to avoid poles. We can pick for example $t=1/2+iy$ and then close at infinity to the left. This gives a residue at $t=0$, but also at negative integers $t$, so at $t=-n$ for $n=0,1,2,\ldots$.  
We have
\begin{equation} \label{res1}
R_{\infty,\mu}(\Theta,1) = \sum_{n=0}^{\infty}  \mathop{\mathrm{Res}}_{t=-n}\,  \pi^{-2t} \mu^{-t} \Gamma(t)\Gamma(1+t) E_{V,1+t}({\mathcal M}) \; .\end{equation}
Let us first consider the $t=0$ pole. Eq.\ \eqref{res1} has a single pole at $t=0$
from $\Gamma(t)$, so taking the residue picks out the {\it constant} term from $E_{V,1+t}$ as $t\rightarrow 0$.
Contrast the massless case when picking out the {\it single pole} in $s$. The constant term contains the ``$\log^2$'' piece
\begin{equation}
R_{\infty,\mu}(\Theta,1)\Big|_{ {\rm log}^2}
 =\mathop{\mathrm{Res}}_{t=0} \,  \pi^{0} \mu^{0} t^{-1}\Gamma(1) E_{V,1+t}({\mathcal M})\Big|_{ {\rm log}^2} ={6 \over \pi}
 \log(\sqrt{T_2}|\eta(T)|^2)\log(\sqrt{U_2}|\eta(U)|^2) \; . 
\end{equation}
This is  not a DKL correction. Instead, it is more like a ``Sudakov'' double logarithm.\footnote{In QED and QCD, a Sudakov double logarithm in momentum
arises when there are two mass scales available \cite{Cohen:2019wxr}.}
Similarly to the Fourier expansion in $\tau_1$, for the remaining poles at negative integers $t=-n$
we use  the  functional relation (reflection
formula) eq.\ \eqref{refl1} for $E_{V,1+t}$ to map it to $1-(1+t)=-t=n$. This produces an additional Gamma function and a zeta function
in numerator and denominator.
We have for the $t<0$ ($n>0$) terms
 \begin{eqnarray}
R_{\infty,\mu}(\Theta,1)\Big|_{t<0}
 &=&\sum_{n=1}^{\infty} \mathop{\mathrm{Res}}_{t=-n}\pi^{-2t-1/2} \mu^{-t}{ \Gamma(-t)\Gamma(t)\Gamma(t+1)\zeta(-2t) \over \Gamma(-t-1/2)\zeta(-2t-1)}{E}_{{\rm V},-t}\nonumber \\
 &=&-\sum_{n=1}^{\infty} \mathop{\mathrm{Res}}_{t=-n}  {2^{-2(t+1)}\pi^{1-2t}\mu^{-t} \over t \sin^2 \pi t }{\zeta(-2t) \over \zeta(-2t-1)\Gamma(-2t-1)}{E}_{{\rm V},-t}\;  . \label{EVres}
 \end{eqnarray}
In the last line, we used the functional relation for $\zeta$. This is like a ``reflected'' version of the smooth $\mu=1$ expansion \eqref{mu1}. Explicitly,
 \begin{eqnarray}
R_{\infty,\mu}(\Theta,1)\Big|_{t<0}
 &=& {\pi^3 \over 3}\mu E_{\rm V,1} + {\pi^3 \over 180\zeta(3)^2}  \mu^2 (c_2 E_{{\rm V},2} + c_2'E'_{{\rm V},2})
 + {\pi^5 \over 8505 \zeta(5)}\mu^3  (c_3 E_{{\rm V},3} + c_3'E'_{{\rm V},3}) + \ldots  \nonumber
  \end{eqnarray}
  where
 \begin{eqnarray}
c_1 &=&   -4 \pi^4 \zeta'(3)+360 \zeta(3) \zeta'(4)+\pi^4 \zeta(3) (2 \log(\mu)+4 \gamma-7+4 \log(\pi)+\log(16)) \nonumber \\
c_1' &=&  2\pi^4 \zeta(3) \\
c_2 &=&   -4 \pi^6 \zeta'(5)+3780 \zeta(5) \zeta'(6)+\pi^6 \zeta(5) (2 \log(\mu)+4 \gamma-9+4 \log(\pi)+\log(16))  \nonumber \\
c_2' &=&  2\pi^6 \zeta(5) \nonumber 
 \end{eqnarray}
The first term with $E_{{\rm V},1}$ is shorthand to use Kronecker's 1st limit formula,
which gives a finite result for the order $\mu$ term that we computed, but we do not display the expression here. 

The total massive theta lift is the sum of the two kinds of contribution:
\begin{equation}
R_{\infty,\mu}(\Theta,1) =  R_{\infty,\mu}(\Theta,1)\Big|_{t=0} + R_{\infty,\mu}(\Theta,1)\Big|_{t<0} \; . 
\end{equation}
where the $t=0$ term includes the log${}^2$ term, and the $t<0$ terms the ``tail'' of higher $E_{V,n}$.
This tail is typical for the massive theta lift, and can be thought of as in the smooth $\mu=1$ expansion \eqref{mu1}. 
The residue \eqref{EVres} gives an explicit way to compute it without using known properties of Meijer G functions (Mellin-Barnes representations).

Another way to put it is that contour integration calculations like those in this section show some properties of those representations.  

\subsection{Alternative: cutoff}
\label{alternative}

This section provides an alternative to the main computations above, and can be skipped.

In the previous sections, we   defined analytic continuations by contour deformation in the complex plane, which is built into the Meijer G (Mellin-Barnes) representation. 
An alternative to contour deformation is to define the analytic continuation by a cutoff procedure. In this section, we will  illustrate this at the simplest level of a double sum $\Theta_2$ as opposed to the quadruple sum $\Theta_{2,2}$, but the principle is the same.

In appendix \ref{sec:analytic}, we review the standard argument that gives analytic continuation of the Eisenstein series $E_s(U)$ as integral representation,
following the logic in Chapter 1 of Bump's book \cite{bump}. 
The final result \eqref{analytic} in the appendix is an explicit analytic continuation, that we repeat here for convenience:
\be   \label{analyticHere}
\int_0^{\infty}\tau_2^{s-1}   \Theta_2(U, \tau_2)
d\tau_2 = \!\!\sum^{\quad \prime}_{n_1,n_2}\!\! {\tt E}_{s}\left({\pi \over U_2}| n_2 U -n_1|^2\right) +{1\over s-1}
+\!\!\sum^{\quad \prime}_{n_1,n_2}\!\! {\tt E}_{1-s}\left({\pi \over U_2}| n_2 U -n_1|^2\right) -{1\over s} \;. 
\ee
where ${\tt E}_{s}$ is the exponential integral
\be
{\tt E}_s(z) = \int_1^{\infty} t^{-s} e^{-zt} dt
\ee
that is closely related to the incomplete gamma function, but this form suits our purposes better.

In this section, we mass deform eq.\ \eqref{analyticHere}:
\be
\underbrace{
2\mu^{s/2}{\pi^s \over \Gamma(s)} \int_0^{\infty}\tau_2^{s/2-1}K_{s}\left(2\pi\sqrt{\mu \over \tau_2}\right)
 \!\!\sum^{\quad \prime}_{n_1,n_2}\!\!e^{-  {\pi\tau_2 \over  T_2 U_2}| n_2 U -n_1|^2} d\tau_2}_{{\tt E}_{s,\mu}[T,U]} +\underbrace{2\mu^{s/2}{\pi^s \over \Gamma(s)} \int_0^{\infty}\tau_2^{s/2-1}K_{s}\left(2\pi\sqrt{\mu \over \tau_2}\right)d\tau_2}_{``{\rm massive\,  pole"}}  
 \ee
 where the summand in the second line is the massive analog of the exponential integral ${\tt E}_s$:
 \be 
\sum^{\quad \prime}_{n_1,n_2} {\tt E}_{s,\mu}(\pi| n_2 U -n_1|^2/(U_2 T_2)) = 
2\mu^{s/2} \!\!\sum^{\quad \prime}_{n_1,n_2}\!\! {T_2^{s/2} U_2^{s/2} \over | n_2 U -n_1|^s}\int_0^{\infty}\hat{\tau}_2^{s/2-1}K_{s}\left(2\pi\sqrt{\mu \over \hat{\tau}_2T_2 U_2}| n_2 U -n_1|\right)e^{-  \pi\hat{\tau}_2} d\hat{\tau}_2
 \ee
 where we  rescaled the exponent to $-\pi \hat{\tau}_2$.
The double sum is a certain massive Eisenstein series of $U$, but it is not the Poincar\'e series we consider here,
where  the double sum is restricted to coprime integers. 
As above, we will instead perform the integral in each summand, so we obtain a sum over Meijer G functions.

The key point here is the following. In the appendix, since we are free to change variable of integration $\tau_2$ to whatever $\hat{\tau}_2$  is in the exponent, the exponential suppression of the double sum ``slipped out'' of the integrand. Here, the Bessel function prevents this ``slipping out'', so unlike the exponential in the ordinary massless integral, convergence is improved. Note that we are not using any property of the Bessel function other than its exponential suppression for large arguments.

The integral labelled ``massive pole'' is the massive analog of the pole term $1/s$. This can easily be calculated explicitly.
(We only display this for illustration --- we will argue in a moment that this expression is not needed.)
\begin{eqnarray}
&&\hspace{-1cm} 2\mu^{s/2}{\pi^s \over \Gamma(s)} \int_0^{L}\tau_2^{s/2-1}K_{s}\left(2\pi\sqrt{\mu \over \tau_2}\right)d\tau_2 
=L + (\pi^2(2\gamma-1+2\log\pi)\mu+\pi^2\mu\log\mu)\log L - {\pi^2 \over 2}\mu \log^2 L
\nonumber \\
&& -{\pi^ 2\over 6}\mu(3\log \mu(\log\mu+4\gamma-2+4\log \pi)+\pi^2+6+12((\gamma-1)\gamma+\log^2\pi + 2\gamma \log\pi)-12\log\pi)\nonumber.
\end{eqnarray}
Like for the pole term in the massless case, we can subtract this counterterm, so its explicit form is not needed when comparing to the finite value in the sum over the Meijer G function. 

Computing the  sum over the the Meijer G function to high truncation is somewhat time-consuming.
But we can replace the Meijer G function with the asymptotic expression $\sim z^{-2/3} e^{-z^{1/3}}$ to good accuracy for ${\mathcal M}^2\gtrsim 10$. 
A compromise is to use the exact expression up
to $|n_1|,|n_2|,|k_1|,|k_2| \leq 5$ (a few hundred terms), and then the asymptotic expression, with which 
$|n_1|,|n_2|,|k_1|,|k_2| \leq 30$ (about a million terms) is easy to achieve, if desired. Since we are not interested
in high accuracy, we will content ourselves with $|n_1|,|n_2|,|k_1|,|k_2| \leq 5$ and only use the asymptotic expression  as a check.

Here is a more direct comparison to the $[1,\infty]$ cutoff integrals in the appendix. 
We introduce a cutoff Meijer G:
\be \label{Gcutoff}
G_{s,{\rm cutoff}}(X) =2\mu^{s/2} {\pi^s \over \Gamma(s)} \int_1^{\infty} \!\!\tau_2^{s/2-1} K_s\left(2\pi\sqrt{\mu \over \tau_2}\right)e^{-\pi \tau_2 X}d\tau_2
\ee
For $\mu\rightarrow 0$ and $s=1$, this becomes the special case ${\tt E}_0$ of the exponential integral, 
which is elementary:
\be
G_{\rm cutoff}(X)\Big|_{s=1} \; \stackrel{\mu\rightarrow 0}{\rightarrow} \; {e^{-\pi X} \over \pi X}
\ee
Although $G_{s,{\rm cutoff}}$ in eq.\ \eqref{Gcutoff} is explicit enough to compute with, as we now show, it is not a well-known special function. For our purposes we prefer the usual Meijer G function in the preceding (and following) sections.

With our normalization, the factor $\pi^s/\Gamma(s)$ sets the term that was ${\tt E}_1$ in
the appendix to zero at $s=0$. We only need to compute the second term. 
For illustration, at $U=i$ we have
\be  \label{masslessE0}
\sum^{\quad \prime}_{n_1,n_2}E_0\left({|n_2 i-n_1|^2}\right)\Big|_{|n_1|\leq 2, |n_2|\leq 2} =
-1+4{e^{-\pi} \over \pi} +2{e^{-2\pi} \over \pi}+{e^{-4\pi} \over \pi}+8{e^{-5\pi} \over 5\pi}
+{e^{-8\pi} \over 2\pi} = -0.943788
\ee
whereas at the same truncation, we have for the massive version the
values in the table. 
\be
\sum^{\quad \prime}_{n_1,n_2}G_{\rm cutoff,\mu}\left({|n_2 i-n_1|^2}\right)\Big|_{|n_1|\leq 2, |n_2|\leq 2} = \begin{array}{|c|c|} \hline
\mu & \mbox{value} \\ \hline \hline
10^{-1} & -0.980804   \\ \hline
10^{-2} & -0.954947   \\ \hline
10^{-3} & -0.945860   \\ \hline
10^{-4} & -0.944096   \\ \hline
10^{-5} & -0.943829   \\ \hline
10^{-6} & -0.943793   \\ \hline
\end{array}
\ee
For small $\mu$ we approach the massless cutoff value \eqref{masslessE0}.
\begin{figure}[h]
\begin{center}
\includegraphics[width=0.35\textwidth]{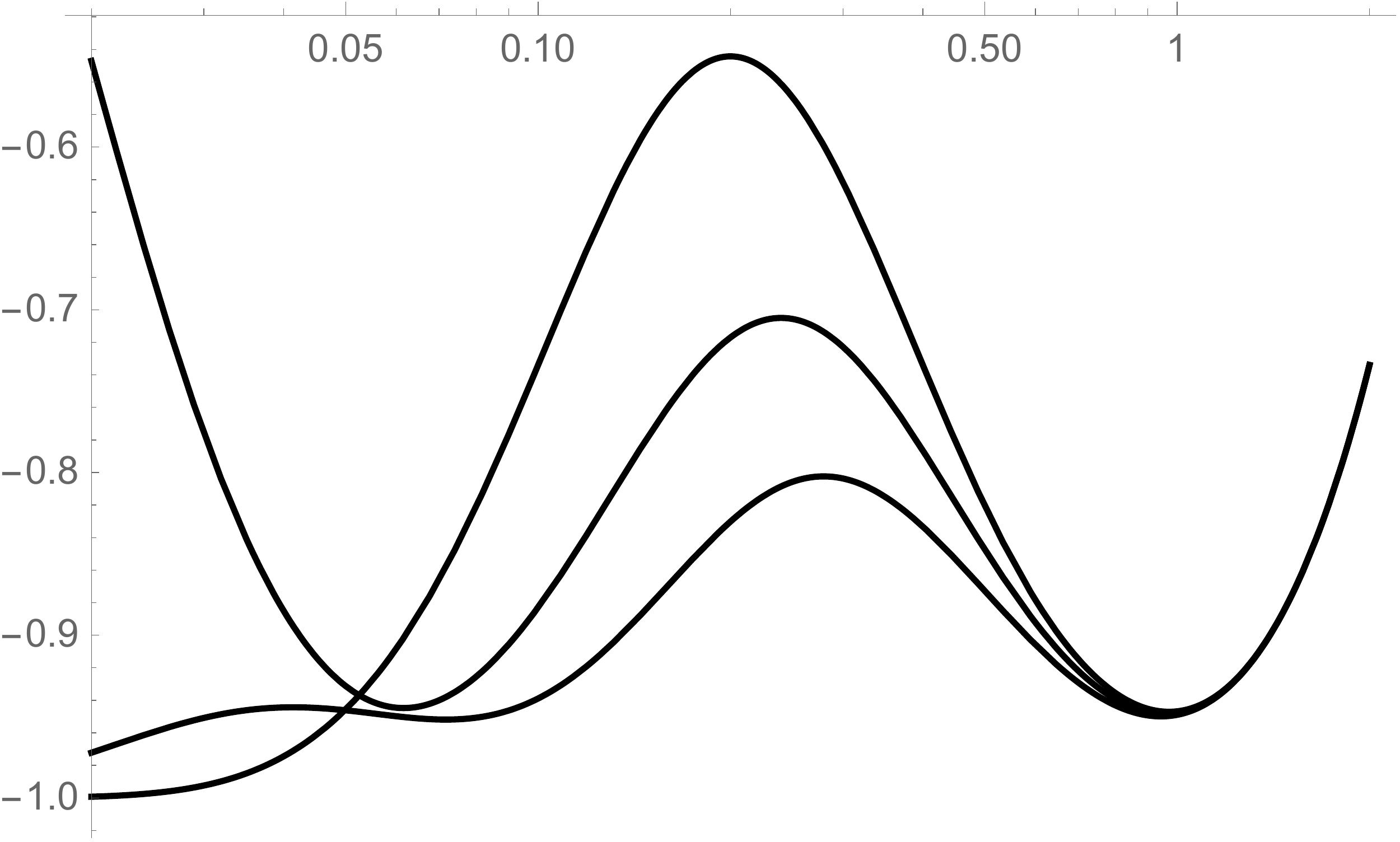}\includegraphics[width=0.35\textwidth]{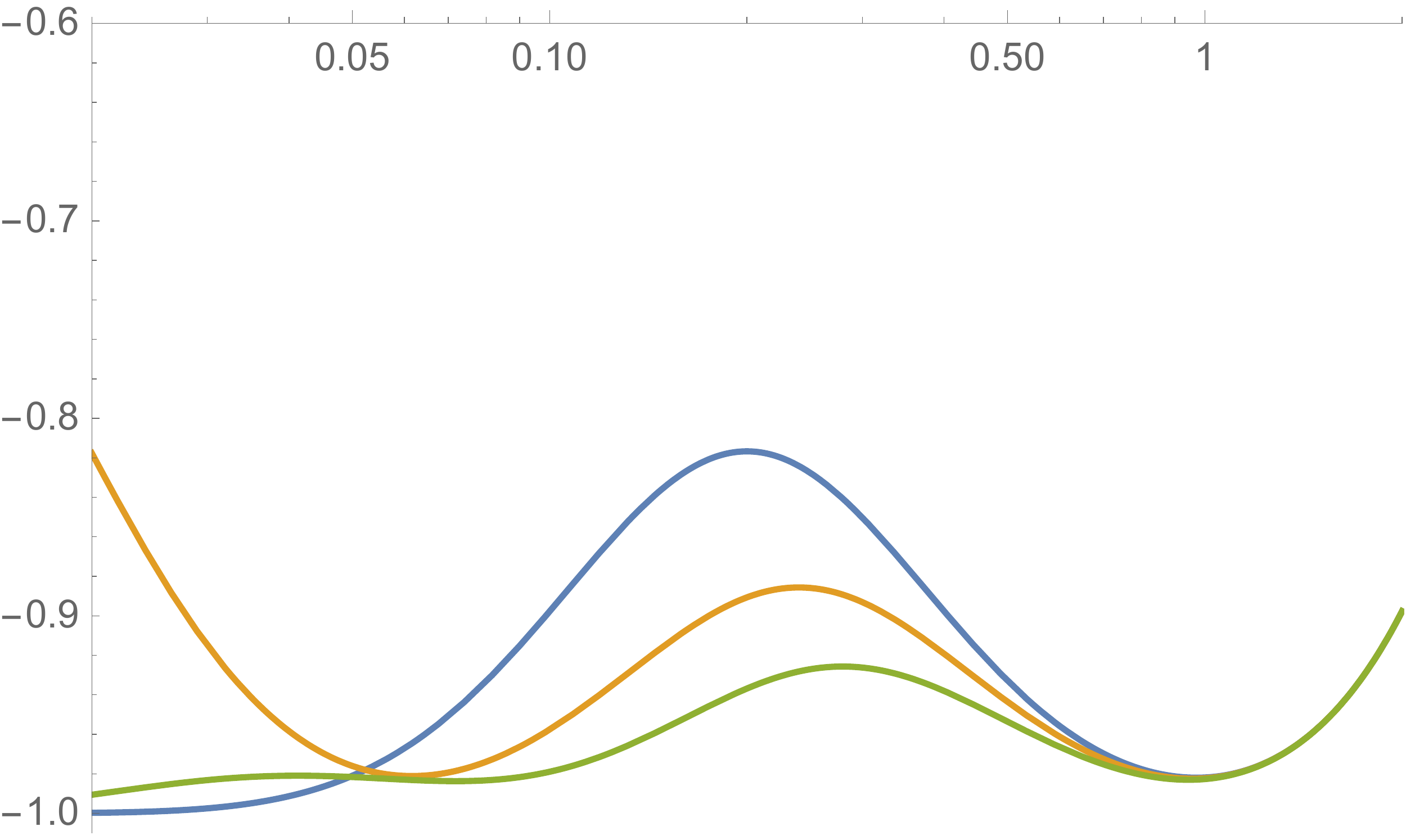}
\caption{Left panel: same three $U_1$ values as in the massless
cutoff in fig.\ \ref{double}, but with our normalization convention.
Right panel: Deformation to $\mu=0.1$, same three $U_1$ values as in previous plot.}
\label{nondeg2}
\end{center}
\end{figure}
We give an illustrative plot in the figure.

\section{Massive theta lift of the $j$-function}
\label{liftingjfunction}
We will now calculate the massive theta lift of the modular invariant $j$-function. The standard case of the theta lift of the $j$-function was reviewed in section \ref{jfunction}, and our result may be viewed as a massive deformation of that result.
\subsection{The theta lift}
The integral we wish to consider is (for later use, we leave $c(0)$ arbitrary)
\begin{equation}
\int_{\mathcal F}j(\tau)\Theta_{2,2}(\tau){\mathcal E}_{s,\mu}(\tau) \, {d \tau d\bar{\tau} \over \tau_2^2}.
\end{equation}
Without the insertion of the massive Eisenstein serie ${\mathcal E}_{s,\mu}(\tau)$ this is the integral of Harvey-Moore. Here with the massive theta lift, we expect not to have to regularize, since as in the massive theta lift without $j$, the Bessel-K and hence Meijer-G function provide exponential suppression for the quadruple sum.
\begin{eqnarray} \label{RSZmassiveT}
R_{L,\mu}(\Theta,s) &=& \int_{\mathcal F_{L}}j(\tau)\Theta_{2,2}(\tau){\mathcal E}_{s,\mu}(\tau) \, {d \tau d\bar{\tau} \over \tau_2^2}\nonumber \\
&=& {\pi^{s}\over \Gamma(s)} \int_{\mathcal S_L^{\rm h}} j(\tau)\Theta_{2,2}(\tau) 2(\mu\tau_2)^{s/2}K_s\left(2\pi \sqrt{\mu/\tau_2}\right)  {d \tau_1d\tau_2 \over \tau_2^2}.
 \end{eqnarray}
We take $L\rightarrow \infty$ to find:
\begin{eqnarray} 
R_{\infty,\mu}(\Theta,s) &=&{2\pi^{s} \mu^{s/2}\over \Gamma(s)}  \!\!\int_0^{\infty} c(m)e^{-2\pi m \tau_2}  \tau_2 \sum_{k\in {\mathbb Z}^{2*}\!,n\in {\mathbb Z}^{2} \atop \langle k,n \rangle =-m}
e^{-{\pi  \tau_2}{\mathcal M}}    \tau_2^{s/2-2}K_s\left(2\pi \sqrt{\mu/\tau_2}\right)d\tau_2 \nonumber \\
&=&\sum_{m=-1}^{\infty} c(m)\!\! \sum_{k\in {\mathbb Z}^{2*}\!,n\in {\mathbb Z}^{2} \atop \langle k,n \rangle =-m}\!\! {2\pi^{s} \mu^{s/2}\over \Gamma(s)} \!\!\int_0^{\infty} 
 \tau_2^{s/2-1} e^{-{\pi  \tau_2}({\mathcal M}+2m)}   K_s\left(2\pi \sqrt{\mu/\tau_2}\right)d\tau_2. 
 \end{eqnarray}
 At this point, one can be worried about the $1/q$ pole ($m=-1$ term) due to the $j$ function which gives rise
 to the ${\mathcal M}+2m$ shift, which for $m=-1$ is ${\mathcal M}-2$. However, 
 the minimum value of ${\mathcal M}$ is 2, so  the Siegel-Narain $\Theta$ takes care of the $\tau_2\rightarrow \infty$ behavior.
The massless theta lift from the previous section appears in each
 term in the $m$ sum:
  \begin{eqnarray}
R_{\infty,\mu}(\Theta,s)  &=& \sum_{m=-1}^{\infty} c(m) \!\!  \sum_{k\in {\mathbb Z}^{2*}\!,n\in {\mathbb Z}^{2} \atop \langle k,n \rangle =-m} \!\! {2\pi^{s} \mu^{s/2}\over \Gamma(s)} \cdot  {\pi^s \over 2}\mu^{s/2}
 G^{3,0}_{0,3}(\pi^3 \mu  ({\mathcal M}+2m)|0,0,-s)\nonumber\\
 &=&  \sum_{m=-1}^{\infty} c(m) {\pi^{2s} \over \Gamma(s)} \sum_{k\in {\mathbb Z}^{2*}\!,n\in {\mathbb Z}^{2} \atop \langle k,n \rangle =-m}\mu^s G^{3,0}_{0,3}(\pi^3 \mu ( {\mathcal M}+2m)|0,0,-s).
\end{eqnarray}

The advantage of a regularization like the above is that we can use the $q$-series for $j(\tau)$ directly. Without regularization,
 the ``naive'' massless $j$-lift is singular, so the above is not directly  a deformation of the naive $j$-lift. Instead, we
 can compare it to the  Selberg-Poincar\'e regularization \eqref{SelbergPoincare} for $s=\epsilon>0$ (where the original holomorphic $j$ is $s=0$), and include massive regularization. 
Writing $\ell$ in the Selberg-Poincar\'e  series, so that the above is the $\ell=1$ special case, we have
\begin{eqnarray} \label{RSZSelbergmassiveT} 
R_{L,\mu}(\Theta,s)(T,U) &=& \int_{\mathcal F_{L}}\underbrace{E^{(\ell)}_{\epsilon}(\tau)}_{\rm unfold}\Theta_{2,2}(T,U,\tau){\mathcal E}_{s,\mu}(\tau) \, {d \tau d\bar{\tau} \over \tau_2^2}\nonumber \\
&=&\int_{\mathcal S_L^{\rm h}} \tau_2^{\epsilon} e^{-2\pi i \ell \tau_1}e^{2\pi  \ell \tau_2}\Theta_{2,2}(T,U,\tau) \sum_m{f}_{s,\mu}^{(m)}(\tau_2)e^{2\pi i m \tau_1} {d \tau_1d\tau_2 \over \tau_2^2} \; . 
 \end{eqnarray}
 For $L\rightarrow \infty$, we have
 \begin{eqnarray}
R_{\infty,\mu}(\Theta,s)(T,U)&=&\sum_m \sum_{k\in {\mathbb Z}^{2*}\!,n\in {\mathbb Z}^{2} \atop \langle k,n \rangle =\ell-m}  \int_0^{\infty} \tau_2^{-1+\epsilon}
e^{-\pi  \tau_2({\mathcal M}(T,U)-2\ell)}  f^{(m)}_{s,\mu}(\tau_2) d\tau_2  
 \end{eqnarray}
 where the $\tau_1$ Fourier coefficients $f^{(m)}_{s,\mu}(\tau_2)$ of  ${\mathcal E}_{s,\mu}(\tau)$ are given in \eqref{fFourier}. 
 Note that the $q$-expansion coefficients of $j$ do not occur explicitly in this expression. So if the two versions
 of the calculation are to agree,  those coefficients must be encoded in the right-hand side. Indeed, the Petersson-Rademacher expansions 
 express $c(m)$ in 
 sums over Kloosterman sums and $I$ Bessel functions. One difference is that the Kloosterman sums that occur for the Eisenstein series $E$ always have one argument zero, whereas the Selberg-Poincar\'e representation of the $j$ function has one argument equal to one.
 
We now have two alternative regularizations of the same integral, so the interesting question arises whether they agree for finite $\mu$. 
 To check this, we can consider the simpler case of just $R_{\infty,\mu}(1,s)$ without the $\Theta$:
   \begin{eqnarray}  \label{unf2}
R_{L,\mu}(1,s) &=& \int_{\mathcal F_{L}}j(\tau)\underbrace{{\mathcal E}_{s,\mu}(\tau)}_{\rm unfold} \, {d \tau d\bar{\tau} \over \tau_2^2}
\quad \stackrel{?}{=}\quad \lim_{\epsilon\rightarrow 0}  \int_{\mathcal F_{L}}\underbrace{E^{(\ell)}_{\epsilon}(\tau)}_{\rm unfold}{\mathcal E}_{s,\mu}(\tau) \, {d \tau d\bar{\tau} \over \tau_2^2} \quad (\ell=1) \; . 
\end{eqnarray}
At this point,  it might be of use to compare our calculations to 
the beautiful work on Niebur-Poincar\'e representations in \cite{Angelantonj:2012gw,Angelantonj:2015rxa}. There, Selberg-Poincar\'e sums like $E^{(\ell)}_{\epsilon}(\tau)$ come with $I$ Bessel functions, since they must have $1/q^{\ell}$ poles at $q\rightarrow 0$, and are deformed to Whittaker functions. 
Here, massive Eisenstein series do not have $1/q$ poles,
and are represented with $K$ Bessel functions, and deform the massless Eisenstein series $E_s$ in the RSZ transform to massive ${\mathcal E}_{s,\mu}$, as in \eqref{unf2}. 

We take $L\rightarrow \infty$:
\be
2 \mu^{s/2} c(0)\int_0^{\infty}  \tau_2^{s/2-2}K_s\left(2\pi \sqrt{\mu/\tau_2}\right)d\tau_2 \nonumber
\; \stackrel{?}{=} \lim_{\epsilon\rightarrow 0} \;  \int_0^{\infty} \tau_2^{-1+\epsilon}
e^{2\pi \ell \tau_2}  f^{(\ell)}_{s,\mu}(\tau_2) d\tau_2    \quad (\ell=1)
 \label{c0val}
\ee
where again $f^{(m)}_{s,\mu}(\tau_2)$ is given in \eqref{fFourier}. On the left,
the $q^0$ term of $j(\tau)$ was singled out by the $\tau_1$ integral. On the right, the $\ell$th Fourier mode of ${\mathcal E}_{s,\mu}(\tau)$ was singled out, since
$E^{(\ell)}_{\epsilon}(\tau)$ contributes $e^{-2\pi i  \ell \tau_1}$, so $e^{2\pi i m \tau_1}$ only gives nonzero $\tau_1$ integral for $m=\ell$. 
For $j(\tau)$, we have $\ell=1$. 

The relation \eqref{c0val} effectively fixes the value of $c(0)$. Of course, $j(\tau)$ is modular invariant irrespective of the value of $c(0)$. But Selberg-Poincar\'e series regularization of $j(\tau)$ fixes $c(0)=12$ in one convention. So it is not surprising that there is also a value fixed here.

\subsection{Mellin-dual massive $j$-theta lift}
As in section \ref{mellindual}, since we find that the massive theta lift is naturally a Mellin integral representation, we could
also have compute the Mellin forward transformation $\mu\rightarrow t$ directly:
\begin{eqnarray} \label{RSZmassiveTM}
{\mathbf M}(R_{\infty,\mu}(\Theta,s))(t) &=& \int_0^{\infty} \mu^{t-1} \left(\int_{\mathcal F_{L}}j(\tau)\Theta_{2,2}(\tau){\mathcal E}_{s,\mu}(\tau) \, {d \tau d\bar{\tau} \over \tau_2^2}\right) d\mu \nonumber \\
&=& \int_{\mathcal S} j(\tau)\Theta_{2,2}(\tau) \pi^{-2t}{\Gamma(t)\Gamma(s+t) \over \Gamma(s)} \tau_2^{s+t-2}  {d \tau_1d\tau_2 \over \tau_2^2}\nonumber \\
&=&\sum_{\ell}c(\ell)\pi^{-2t}{\Gamma(t)\Gamma(s+t) \over \Gamma(s)} \sum_{k\in {\mathbb Z}^{2*}\!,n\in {\mathbb Z}^{2} \atop \langle k,n \rangle =-\ell} \int_{\mathcal S} \tau_2 
e^{-{\pi  \tau_2}({\mathcal M}+2\ell)}  \tau_2^{s+t-2}  {d \tau_1d\tau_2 \over \tau_2^2}\nonumber \\
&=&\sum_{\ell}c(\ell)\pi^{-s-3t} {\Gamma(t)\Gamma(s+t)^2 \over \Gamma(s)}\sum_{k\in {\mathbb Z}^{2*}\!,n\in {\mathbb Z}^{2} \atop \langle k,n \rangle =-\ell} ({\mathcal M}+2\ell)^{-(s+t)}\nonumber\\
&=&\sum_{\ell}c(\ell) \pi^{-2t} {\Gamma(t)\Gamma(s+t) \over \Gamma(s)} E^{(\ell)}_{V,s+t} \label{res}
\end{eqnarray}
where $E^{(\ell)}_{V}$ is a shifted (${\mathcal M}\rightarrow {\mathcal M}+2\ell$) non-degenerately constrained ($\langle k,n\rangle=0 \rightarrow \langle k,n\rangle=-\ell$) lattice Eisenstein series. Apart from that, each term is exactly like without the $j$ function.
As before, we can fix $s=1$ and pick out poles for $t=-n$ with $n=0,1,2,\ldots$.
\begin{equation}  \label{RSZmassiveTM}
R_{\infty,\mu}(j\Theta,1) =\sum_{\ell}c(\ell)  \sum_{n=0}^{\infty}  \mathop{\mathrm{Res}}_{t=-n}\,  \pi^{-2t} \mu^{-t} \Gamma(t)\Gamma(1+t) E^{(\ell)}_{V,1+t} \; . \end{equation}
The drawback with this representation in this context
is that the constrained sums give no indication of how 
differential operators act on this composite object. On the 
other hand, we see in   \cite{Angelantonj:2011br} that the sum with
constraint $\langle k,n\rangle=-\ell$ is found from the first constrained sum $\ell=1$ from acting with Hecke
operators.

\section{Conclusions and Outlook}
\label{conclusions}
In this paper we have taken the first steps toward analyzing theta lifts of massive modular forms. Our work points to several interesting questions that we leave for future research. Below we provide a brief summary.

We have obtained explicit expressions for theta lifts, but a satisfactory interpretation of this is still lacking. As mentioned in the introduction, a key aspect of the theta lift is to transfer automorphic forms (or automorphic representations) from one Lie group to another. This fact plays an important role also in string theory, where the amount of supersymmetry dictates the relevant automorphic representation. It is therefore natural to wonder if there is a corresponding representation-theoretic statement underlying the massive theta lift. In physics,  non-perturbative solutions (solitons) are known to correspond to certain massive (BPS) representations of the (super-)Poincar\'e algebra. 
Plane-wave limits of any solution
from a perturbative viewpoint corresponds to fields that satisfy a harmonic oscillator differential equation in spacetime  \cite{Blau}. For maximal supersymmetry, these BPS-representations are intimately related to small automorphic representations (see e.g. \cite{Fleig:2015vky}). It would be very interesting to understand whether there exists  corresponding massive automorphic representations underlying the image of the massive theta lift.

In this paper we have treated the mass $\mu$ as an external parameter, i.e. the massive modular forms really correspond to families of modular forms parameterized by $\mu\in \mathbb{R}_{\geq 0}$. It would be interesting to investigate whether it is possible embed this construction into a larger Lie group, where $\mu$ would be promoted to an honest modulus. One natural conjecture is that the massive Eisenstein series occur as Fourier coefficients of some automorphic form on $SL(3,\mathbb{R})$. For example, it would be very interesting to understand whether there is a relation to very similar formulas in \cite{Vinogradov,BumpChoie,Pioline:2009qt}.

We have considered theta lifts from $SL(2,\mathbb{R})$ to $SO(d,d)$. The theta correspondence works however in far greater generality, and allows for transferring automorphic representations between a wide class of groups, known as {\it dual reductive pairs}. For example, in the context of the pair $SL(2,\mathbb{R})\times SO(d,d)$ one can also consider theta ``lifts'' from $SO(d,d)$ to $SL(2,\mathbb{R})$. In practise this involves integrating over $\Gamma\backslash SO(d,d)/(SO(d)\times SO(d))$ to produce a modular form on $SL(2,\mathbb{R})$. For certain special choices this is well known in mathematics as the {\it Siegel-Weil formula}. This type of integral was also considered recently in string theory by Maloney and Witten  \cite{Maloney:2020nni}, as a way of averaging over Narain moduli space. A similar type of integral  would be straightforward to study in the massive case, using the results in this paper. We hope to return to this question in the future. 

Finally, let us note that calculations such as those in this paper can also apply to condensed matter physics. One application for the kind of integration performed here is that the area of a closed subregion gives the Berry phase in topological systems \cite{Levay:2019nsr}. Another long-standing topic in condensed matter physics is the statistical physics of the Ising model away from criticality, as in recent work by one of us \cite{Marcus}. In fact, methods of this kind were used in statistical physics before they made it to high-energy physics \cite{Saleur1987}.

\section*{Acknowledgments}
We are grateful to Dan Bump, YoungJu Choie, Jens Funke, Joe Minahan, Federico Zerbini and Sol Friedberg for helpful discussions and correspondence. We would also like to thank the Isaac Newton Institute of Mathematical Sciences, Cambridge, for support and hospitality during the programme ``New connections in number theory and physics'' when work on this paper was undertaken.  D.P. was supported by the Swedish Research Council (grant no. \ 2018-04760), and the Wallenberg AI, Autonomous Systems and Software Program (WASP) funded by the Wallenberg Foundation (grant no.\ 2020.0173).

\appendix
\section{Fourier-transforming Bessel functions with square roots}
\label{FourierBessel}
The Fourier integrals (zero mode and nonzero mode) are of the form \cite{Bateman} (1.13.45):
\be \label{FourierK}
\int_0^{\infty} dx  \cos(xy) (x^2+\beta^2)^{-\nu/2}K_{\nu}(\alpha\sqrt{x^2+\beta^2}) = \sqrt{\pi \over 2} \alpha^{-\nu}\beta^{1/2-\nu}(y^2+\alpha^2)^{\nu/2-1/4} K_{\nu-1/2}(\beta\sqrt{y^2+\alpha^2})
\ee
for Re $\alpha>0$, Re $\beta>0$.

At the time of writing, Mathematica does not perform the integral directly,
but it can do the following two steps. Using the following Mellin representation
\be
 (x^2+\beta^2)^{-\nu/2}K_{\nu}(\alpha\sqrt{x^2+\beta^2})  = \int_0^{\infty} \! dt \; 2^{-\nu-1}\alpha^{-\nu-1}e^{-{\alpha^2(x^2+\beta^2) \over 4t}-t} \; , 
\ee
we can easily Fourier transform the Mellin integrand:
\be
\int_0^{\infty}\! dx\, \cos(xy)\,  2^{-\nu-1}\alpha^{-\nu-1}e^{-{\alpha^2(x^2+\beta^2) \over 4t}-t} = \sqrt{\pi}\, 2^{-\nu-1}\alpha^{\nu-1}t^{-\nu-1/2}e^{{\alpha^2 \beta^2 \over 4t}
-{t y^2 \over \alpha^2}-t} \; . 
\ee
Now undoing the Mellin representation by performing the $t$ integral gives eq.\ \eqref{FourierK} above,
where for our purposes it is sufficient to make the stronger assumption $\alpha>0$, $\beta>0$, $\nu>0$.

\section{Massive deformations of  holomorphic objects}
\label{holo}

The massive Maass-Jacobi form ${\mathcal E}_{s, \mu}(\tau, z)$ is nonholomorphic,
but there is a sense in which it also provides deformations of holomorphic objects. We do not expect the deformation to preserve holomorphy,
since part of the definition of the mass-deformed objects is that they satisfy other differential equations than the Cauchy-Riemann equations. Let us briefly explain this.

The mass-deformed Maass form ${\mathcal E}_{\mu,s}(\tau)$  descends from the Jacobi-Maass form \cite{Berg:2019jhh}
\be \label{MJ}
{\mathcal E}_{\mu,s}(z;\tau) = 2 \sum^{\quad \prime}_{m,n}\left({\sqrt{\mu \tau_2} \over |m\tau + n|}\right)^{s}K_s\left(2\pi \sqrt{\mu \over \tau_2}|m\tau+n|\right)
e^{{2\pi i \over \tau_2} {\rm Im} ((m\tau+n)\overline{z})} 
\ee
which is a deformation of the massless ($\mu=0$) nonholomorphic Kronecker-Eisenstein series
\be
E_{s}(z;\tau) =  \sum^{\quad \prime}_{m,n}{\tau_2^s\over |m\tau + n|^{2s}}
e^{{2\pi i \over \tau_2} {\rm Im} ((m\tau+n)\overline{z})} =\sum^{\quad \prime}_{m,n}{\tau_2^s\over (m\tau + n)^s(m\bar{\tau} + n)^s}
e^{{\pi  \over \tau_2}((m\tau+n)\overline{z}-(m\bar{\tau}+n)z)}.
\ee
Setting $s=2k \in 2{\mathbb Z}$, differentiating $2k$ times with respect to $z$, we obtain
\be  \label{der}
  \partial^{2k}_z E_{s}(z;\tau) =  \pi^{2k}\sum^{\quad \prime}_{m,n}{1\over (m\tau + n)^{2k}}
e^{{2\pi i \over \tau_2} {\rm Im} ((m\tau+n)\overline{z})} 
\ee
so setting $z=0$, we have the ordinary holomorphic Eisenstein series
\be \label{ordin}
E^{\rm hol}_{2k}(\tau) = {1 \over  \pi^{2k}} \partial^{2k}_z E_{2k}(z;\tau)\Big|_{z=0} =  \sum^{\quad \prime}_{m,n}{1\over (m\tau + n)^{2k}} \; . 
\ee
The ``generating function'' \eqref{der} suggests to define a massive deformation of $E^{\rm hol}_{2k}(\tau)$ by $\mu$ as 
\be \label{Ehol}
{\mathcal E}^{\rm hol}_{\mu,2k}(\tau) = {1 \over  \pi^{2k}} \partial^{2k}_z {\mathcal E}_{\mu,2k}(z;\tau)\Big|_{z=0} 
=2 \sum^{\quad \prime}_{m,n}(\mu \tau_2)^{k}\left({m\overline{\tau} + n \over m \tau + n}\right)^{k}K_{2k}\left(2\pi \sqrt{\mu \over \tau_2}|m\tau+n|\right)
\ee
with the massive Jacobi-Maass form from \eqref{MJ}. In general, this has no reason to be holomorphic, but unlike \eqref{MJ}, 
one specific term in \eqref{Ehol} coming from the series expansion of the Bessel function in $\mu$ will be holomorphic,
and there will formally be a ``tail'' of nonholomorphic contributions, much like the ``shadow'' of Niebur-Poincar\'e series. 
This expansion is only formal, since the double sum of the series representation does not converge. We can either simply avoid series-expanding the Bessel function, or use the integral representation
of \eqref{MJ} instead of the double sum.

Note that while ${\mathcal E}_{\mu,s}(0;\tau)$ has an obvious Poincar\'e series representation, the deformation \eqref{Ehol} of $E_{2k}$
does not have an obvious Poincar\'e series generated purely from $\tau_2$. This means
that the calculation of the Fourier expansion does not apply
and would need to be generalized.  For the purposes of this discussion we can stick to \eqref{Ehol} as
a definition of ${\mathcal E}^{\rm hol}_{\mu,2k}(\tau)$.

From \eqref{Ehol} we can define a massive deformation of the $j$-function
\be
j_{\mu}(\tau) = 1728 {{\mathcal E}^{\rm hol}_{\mu,4}(\tau)^3 \over {\mathcal E}^{\rm hol}_{\mu,4}(\tau)^3-{\mathcal E}^{\rm hol}_{\mu,6}(\tau)^2}
\ee
that evidently by \eqref{ordin} reduces for $\mu=0$ to the usual $j$-function
\be
j_{\mu=0}(\tau) = 1728 {E_4^3(\tau) \over E_4^3(\tau)-E_6^2(\tau)} \; . 
\ee
Of course, when dealing with deformations, we should explore whether this deformation $j_{\mu}(\tau)$ is equivalent, in the sense of \cite{Berg:2019jhh}, 
to other possible deformations of the $j$ function. One possibility for expressing the massive theta lifts
in terms of these types of objects, including also the discriminant function
\be
\Delta_{\mu}(\tau)=
(\eta_{\mu}(\tau))^{24} =  {\mathcal E}^{\rm hol}_{\mu,4}(\tau)^3-{\mathcal E}^{\rm hol}_{\mu,6}(\tau)^2 \; . 
\ee
For basic numerical checks, the Fourier series converges much faster than the double sum. (Alternatively,
the integral representation is also exponentially convergent.)

Incidentally, if we don't set $s=2k$ above, we have
\be \label{Ehol2}
{\mathcal E}^{\rm hol}_{\mu,s,2k}(\tau) = {1 \over  \pi^{2k}} \partial^{2k}_z {\mathcal E}_{\mu,s}(z;\tau)\Big|_{z=0} 
\ee
so for $\mu=0$ we  generate the mixed object
\be
{\mathcal E}^{\rm hol}_{\mu=0,s,2k}(\tau)= \sum^{\quad \prime}_{m,n}{ (m\overline{\tau} + n)^{2k}\over|m\tau+n|^{2s}} =\sum^{\quad \prime}_{m,n}{1\over (m\tau + n)^{2k}|m\tau+n|^{2s-2k}} \; . 
\ee

\section{Integral representation as analytic continuation}
\label{sec:analytic}

In the main text we encounter 
\be  \label{firstint2}
 \int {d^2 \tau \over \tau_2^2} \Theta(T,U,\tau) E_s^{(\ell)}(\tau) = E_{{\rm V},s}^{(\ell)}(T,U) 
\ee

In the degenerate case
$\ell=0$, then $E_{{\rm V},s}^{(0)}(T,U)$ is an $SO(2,2)$ vector representation Langlands-Eisenstein series (see for example eq.\ (4.135) in \cite{Fleig:2015vky}),
hence the ``V'' for vector.  

\subsection*{Degenerate lattice Eisenstein series: $\Theta_2$}

We first review the standard argument that gives analytic continuation of $E_s(U)$ as integral representation.
We follow Siegel's Tata notes \cite{SiegelTata}. Some other references are Appendix E in \cite{Berg:2014ama},  Appendix A  in \cite{Berg:2019jhh} 
and Bump's book \cite{bump} where this is Exercise 1.6.2. (Presumably the book's author would require
more rigor for full credit than the meager offering on display here.)

By scaling $\tau_2/T_2\rightarrow \tau_2$, which moves the constant $T_2^s$ out front, 
\be
\Theta_2(U, \tau_2)= \sum_{n_1,n_2}\!\!e^{-  {\pi\tau_2 \over U_2}| n_2 U -n_1|^2} 
\ee
(as opposed to the quadruple sum $\Theta_{2,2}(T,U,\tau)$) so $\Theta_2(U,\tau_2)=\tau_2^{-1}\Theta_2(U,1/\tau_2)$, analogously to the familiar $\vartheta_3(i\tau_2)
=\tau_2^{-1/2}\vartheta_3(i/\tau_2)$. As in Riemann's paper with the integral representation of $\zeta(s)$, we split up the vertical strip integral into $[0,1]$ and $[1,\infty]$:
\be
\int_0^{\infty}\tau_2^{s-1}   \Theta_2(U, \tau_2)
d\tau_2 =
\underbrace{\int_0^{1}\tau_2^{s-1}
 \Theta_2(U, \tau_2)d\tau_2}_{I_1(s)} + \underbrace{\int_1^{\infty}\tau_2^{s-1}  
 \Theta_2(U, \tau_2)d\tau_2}_{I_2(s)}
 \ee
and we use $\Theta_2(U, \tau_2)=\tau_2^{-1}\Theta_2(U, 1/\tau_2)$ in the $[0,1]$ piece $I_1$ and change variable of integration $\widetilde{\tau}_2 = 1/\tau_2$:
\be
I_1(s)=  \int_0^{1}\tau_2^{s-1}  \tau_2^{-1}\Theta_2(U, 1/\tau_2)d\tau_2 
= \int_1^{\infty}\widetilde{\tau}_2^{2-s} \Theta_2(U, \widetilde{\tau}_2)\widetilde{\tau}_2^{-2} d\widetilde{\tau}_2 
= \int_1^{\infty}\widetilde{\tau}_2^{-s} \Theta_2(U, \widetilde{\tau}_2)d\widetilde{\tau}_2 =I_2(1-s)
\ee
Now both terms are over the interval $[1,\infty]$, and the sums converge exponentially. Separating out
the  term
  $(n_1,n_2)=(0,0)$, for $s\geq 0$ we should put an upper cutoff
 $L$ to be able to integrate it. Denoting the sum without   $(n_1,n_2)=(0,0)$ with a prime, we find:
\be  \label{I2piece}
I_2(s)\Big|_L = 
\int_1^{\infty}\tau_2^{s-1}  
 \!\!\sum^{\quad \prime}_{n_1,n_2}\!\!e^{-  {\pi\tau_2 \over U_2}| n_2 U -n_1|^2} d\tau_2 +\int_1^{L}\tau_2^{s-1}d\tau_2  = \!\!\sum^{\quad \prime}_{n_1,n_2}\!\! {\tt E}_{1-s}\left({\pi \over  U_2}| n_2 U -n_1|^2\right) +{L^s-1\over s}
 \ee
where ${\tt E}_{s}$ is the exponential integral
\be
{\tt E}_s(z) = \int_1^{\infty} t^{-s} e^{-zt} dt
\ee
\begin{figure} \label{nondeg2}
\begin{center}
\includegraphics[width=0.4\textwidth]{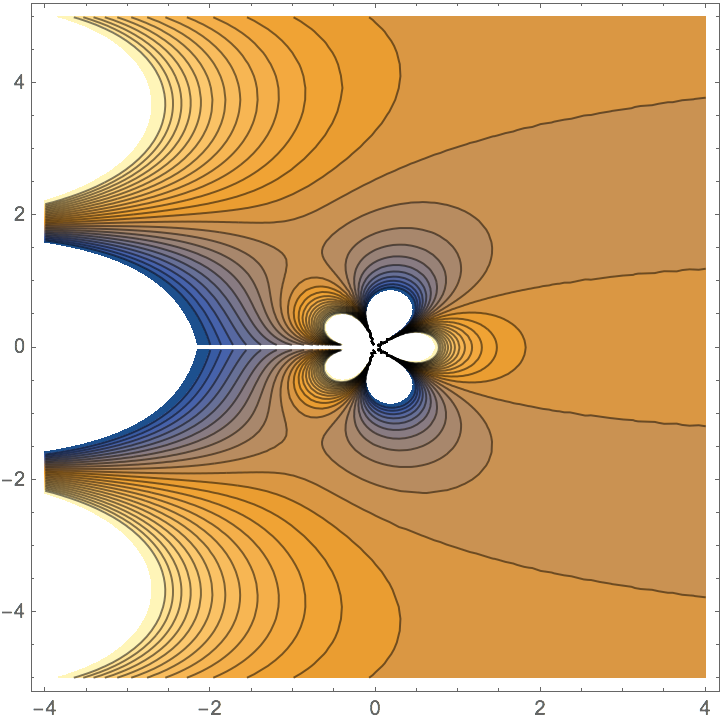}
\end{center}
\caption{Analytic continuation of the exponential integral: Re $E_{-3/2}(z)$.}
\end{figure}
\noindent which for large real arguments $z=x\rightarrow \infty$ is bounded by $e^{-x}\ln(1+1/x)$, so the sum converges quickly. The exponential integral is related to the incomplete gamma function as 
${\tt E}_s(z) = x^{s-1}\Gamma(1-s,z)$, and is a special case of the Kummer $U$ function 
as ${\tt E}_s(z)= e^{-z}U(1,1,z)$. 

We now subtract the term $L^s/s$, take the limit $L\rightarrow \infty$, and the remaining total is
\be   \label{analytic}
\int_0^{\infty}\tau_2^{s-1}   \Theta_2(U, \tau_2)
d\tau_2 = \!\!\sum^{\quad \prime}_{n_1,n_2}\!\! {\tt E}_{s}\left({\pi \over U_2}| n_2 U -n_1|^2\right) +{1\over s-1}
+\!\!\sum^{\quad \prime}_{n_1,n_2}\!\! {\tt E}_{1-s}\left({\pi \over U_2}| n_2 U -n_1|^2\right) -{1\over s} \;. 
\ee
which provides the desired analytic continuation to all $s$ away from the poles at $s=0$ and $s=1$. 
Reflection symmetry $s$ to $1-s$ is manifest from switching the two pairs of terms. 

Contrast this with the direct integration for Re $s>1$:
\be \label{step}
\int_0^{\infty}\tau_2^{s-1}   \Theta_2(U, \tau_2) = \pi^{-s}\Gamma(s)\sum_{n_1,n_2} {U_2^s \over | n_2 U -n_1|^{2s}} 
 =  \pi^{-s}\Gamma(s)2\zeta(2s)E_s(U)=: E_s^{\star}(U) 
\ee
where $E_s^{\star}(U) =
\xi(2s)E_s(U)$ and 
 $\xi(2s) = \pi^{-s}\Gamma(s)\zeta(2s)$. Formally the above observation of the symmetry under $s \leftrightarrow 1-s$ 
verifies that
\be
E_{1-s}^{\star}(U)  = E_{s}^{\star}(U) 
\ee
but of course, the double sum representation is never valid on both sides of this functional relation simultaneously. 
For later, it is useful to give the first equality \eqref{step} in more explicitly: we move
the sum out of the integral, absorb  the exponent's $U$ dependence  into $\tau_2$ 
as $\hat{\tau}_2 ={\tau_2 \over U_2}| n_2 U -n_1|^2 $  and re-emit it
in front:
\be  \label{hatted}
\int_0^{\infty}\tau_2^{s-1}    \sum_{n_1,n_2}\!\!e^{-  {\pi\tau_2 \over U_2}| n_2 U -n_1|^2}\,  d\tau_2
=\sum_{n_1,n_2}\!{U_2^s \over | n_2 U -n_1|^{2s}}\!
\int_0^{\infty}\hat{\tau}_2^{s-1}  e^{-  {\pi\hat{\tau}_2}}\,  d\hat{\tau}_2
\ee
and the remaining $\hat{\tau}_2$ integral produces $\pi^{-s}\Gamma(s)$. Again, this requires Re $s>1$ for the double sum to converge. 

\begin{figure}
\begin{center}
\includegraphics[width=0.6\textwidth]{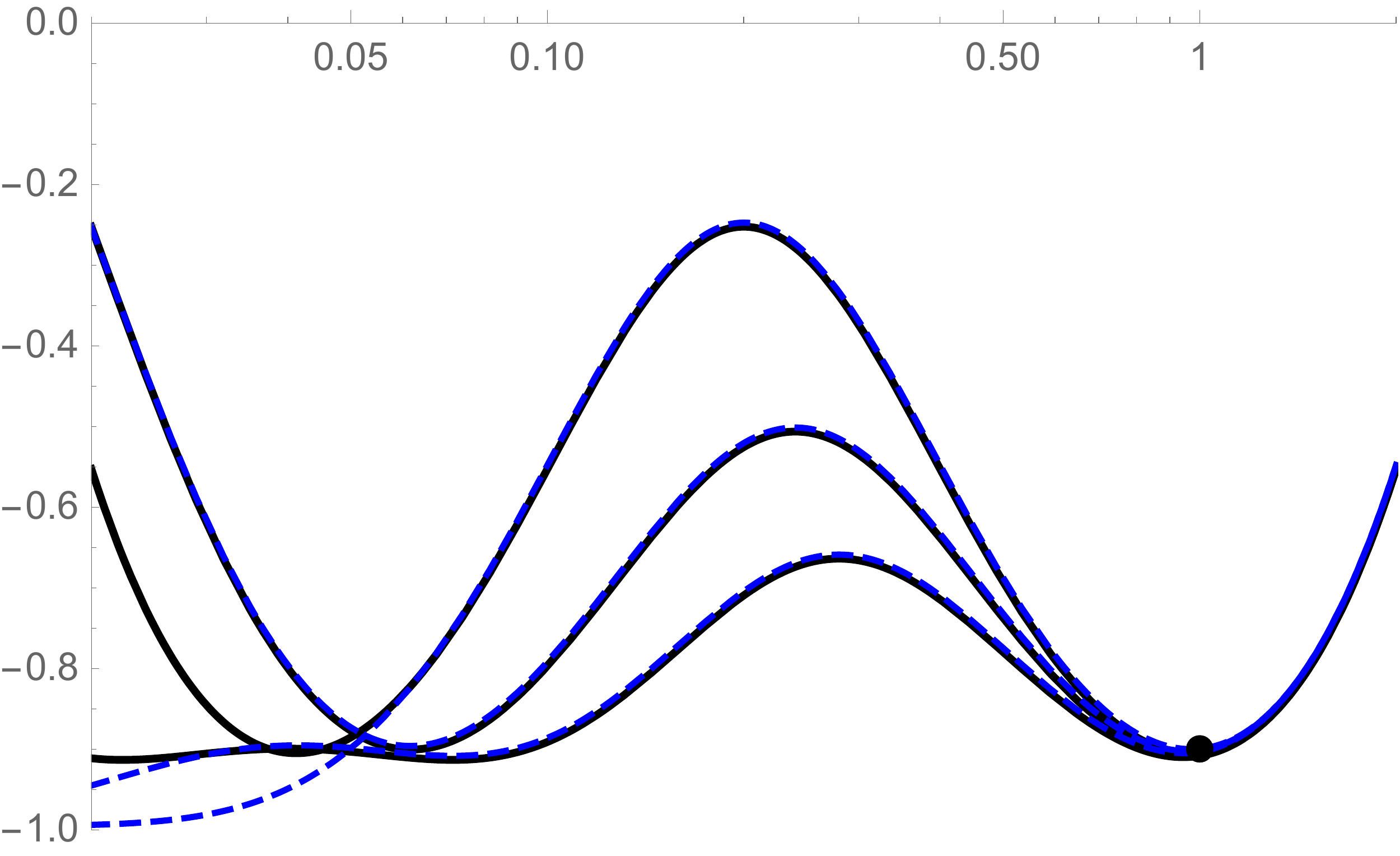}
\vspace{-2mm}
\end{center}
\caption{ \label{double} The truncation eq.\ \eqref{trunc1} to $|n_1|\leq 4, |n_2|\leq 4$ in solid black, and eq.\ \eqref{exact} in dashed blue.
Here $U=0.2,0,24,0.28$. The black dot is $-0.899$ at $U_2=1$. For $|n_1|\leq 5, |n_2|\leq 5$ the curves become indistinguishable.}
\end{figure}

As a concrete example that the double sums
in \eqref{analytic} above provide a sensible truncation prescription, we want to compare to
Kronecker's first limit formula (for more on normalizations, see appendix \ref{norm})
\be   \label{KL1}
E_s(\tau) = {3 \over \pi(s-1)} + {6 \over \pi}\left(12 \log A -\log(4\pi) - \log(\sqrt{\tau_2}|\eta(\tau)|^2)\right) + \ldots
\ee
Let us brutally truncate to $|n_1|\leq 1, |n_2|\leq 1$. We find at $U=i$ that
\be  \label{trunc1}
\left(\mbox{RHS of  eq. } \eqref{analytic} -{1 \over s-1} \right)_{|n_1|, |n_2|\leq 1}=
4({\tt E}_1(\pi) + {\tt E}_1(2\pi)) + {4\over \pi}(e^{-\pi} + {1 \over 2} e^{-2\pi})-1 = -0.89912...
\ee
whereas \eqref{KL1} gives, using $\eta(i) = \Gamma(1/4)^4/(16 \pi^3)$, the exact result
\be  \label{exact}
\left(E_s(i) -{3 \over \pi(s-1)}\right)_{s\rightarrow 1}= \gamma+ \log(4\pi)-4 \log \Gamma(1/4)= -0.89912...
\ee
so agreement is good for such a simple truncation. We plot them for a few values of $U$ in figure \ref{double}. 

\subsection*{Degenerate lattice Eisenstein series: $\Theta_{2,2}$}
\label{analytic4}
For the quadruple sum with $\ell=0$, there one factor in the exponent for $U$ and one factor for $T$,
and only a single variable of integration $\tau_2$, so the summand appears coupled as in \eqref{eqET2}:
\be \label{quadruple}
 \int_0^{\infty}\tau_2^{s-1}\Theta^{(\ell=0)}_{2,2}(T, U, \tau_2)= \int_0^{\infty}\tau_2^{s-1}
 \!\sum_{c,d}\sum_{n_1^*,n_2^*}\!\!e^{-  {\pi\tau_2 \over T_2 U_2}| n_2^* U -n_1^*|^2| d T -c|^2}
 d\tau_2 \; . 
 \ee
For Re $s>1$, the point of \eqref{hatted} was
to make it evident that they both just come out front, which was
\eqref{EVfactor}. We see that it is because there is just a power $\tau_2^{s-1} $ that allows it to factor. (In the massive case, there is not just a power and
it will not factor.) 
\be
E_s^{\star}(T) E_s^{\star}(U)  =  \left({\tt E}_{s}[T] +{1 \over s-1} + {\tt E}_{1-s}[T] -{1 \over s}\right)
\left({\tt E}_{s}[U] +{1 \over s-1} + {\tt E}_{1-s}[U] -{1 \over s}\right)
\ee
with the shorthand ${\tt E}_s[U] = \sum^{\prime}{\tt E}_s(\pi |n_2 U-n_1|^2/U_2)$. So the completed integral
over the quadruple sum must split into these 16 terms.

\section{Contour integrals and the Meijer G function}
\label{Meijer}
Contour integration is the first way that Riemann in his 1859 paper\footnote{For a discussion of Riemann's paper, see \cite{Edwards}.} provided an analytic continuation of $\zeta(s)$, the second way being the type of
analytic continuation we discussed in a previous appendix, with $\vartheta$ in place of $\Theta$.  So it is natural to consider the first point of view here as well.

The Meijer G function is defined from a Mellin-Barnes representation \cite{BatemanHTF}
\be \label{def1}
G^{m,n}_{p,q}\!\left(\begin{array}{c}{\bf a}_p \\ {\bf b}_q \end{array},z,r\right) =
{r \over 2\pi i }\int_C{\prod_{j=1}^m\Gamma(b_j - rs)\prod_{j=1}^n\Gamma(1-a_j + rs) \over \prod_{j=m+1}^q\Gamma(1-b_j + rs) \prod_{j=n+1}^p\Gamma(a_j - rs)}z^s ds
\ee
where ${\bf a}_p = (a_1,\ldots,a_p)$, ${\bf b}_q = (b_1,\ldots,b_q)$
and the contour $C$ separates the poles of $\Gamma(b_1-rs)$, $\ldots \Gamma(b_1-rs)$ 
from poles of $\Gamma(1-a_1+rs)$, $\ldots \Gamma(1-a_n+rs)$. We give an example below. Note that
$q$ and $p$ are in reverse order, and that the original references consider the case $r=1$. When
we omit $r$, we intend $r=1$.\footnote{Mathematica uses $r\neq 1$ for its most basic functions, for example try {\tt MeijerGReduce[BesselK[s,z],z]} which gives $r=1/2$ as default. A representation of $K_s(z)$ with $r=1$
is given in \eqref{rep1}. Also, Mathematica's $s$ is minus that in equation \eqref{def1}, which 
does not change the result  since $s$ is a variable of integration, but it changes the explicit representation of the contour in the sense of orientation, including whether a contour is ``to the left'' or ``to the right'' of poles.} 

In view of the definition  \eqref{def1}, it is not surprising that the Mellin transform of the Meijer G function is simply gamma functions of the parameters and a power:
\be  \label{def2}
\int_0^{\infty} z^{s-1}G^{m,n}_{p,q}\!\left(\begin{array}{c}{\bf a}_p \\ {\bf b}_q \end{array},wz\right) dz=
{\prod_{j=1}^m\Gamma(b_j + s)\prod_{j=1}^n\Gamma(1-a_j - s) \over \prod_{j=m+1}^q\Gamma(1-b_j - s) \prod_{j=n+1}^p\Gamma(a_j + s)}w^{-s}
\ee
A physicist might view the integration along the real axis \eqref{def2} as a more natural starting point for a definition
than  \eqref{def1}, but the real axis is no less arbitrary as a contour than 
 the Mellin contour above.

The set of Meijer G functions is closed under convolution:
\be   \label{conv}
\int_0^{\infty} 
G^{m',n'}_{p',q'}\left(\begin{array}{c}{\bf a}_{p'} \\ {\bf b}_{q'} \end{array},yz\right)G^{m,n}_{p,q}\left(\begin{array}{c}{\bf a}_p \\ {\bf b}_q \end{array},xz\right) dz
 =
G^{m+n',n+m'}_{p+q',q+p'}\!\left(\!\!\!\begin{array}{c}-b_{1'},\ldots, -b_{m'},{\bf a}_p,-b_{m'+1},\ldots -b_{q'} \\ 
-a_{1'},\ldots,-a_{n'},{\bf b}_q,-a_{n'+1},\ldots,-a_{p'} \end{array},{x \over y}\right)
\ee
The $x/y$ introduces an asymmetry, but there is an equivalent representation with $y/x$
using the inversion property
\be \label{inv}
G^{m,n}_{p,q}\left(\begin{array}{c}{\bf a}_p \\ {\bf b}_q \end{array}\Big|z\right) =G^{n,m}_{q,p}\left(\begin{array}{c}1-{\bf b}_q \\ 1-{\bf a}_p \end{array}\Big|z^{-1}\right) 
\ee
One advantage compared to a hypergeometric function is that inverting the argument is still a single Meijer-G.

Representations of known functions include
\be  \label{rep1}
G^{1,0}_{0,1}\left(\begin{array}{c}- \\ b_1 \end{array}, z \right) = z^{b_1} e^{-z} 
\; , \; G^{0,1}_{1,0}\left(\begin{array}{c} a_1 \\ - \end{array}, z \right) = z^{a_1-1} e^{-1/z} \; , \;
G^{2,0}_{0,2}\left(\begin{array}{c} - \\ {b_1,b_2} \end{array} , z \right) = 2z^{(b_1+b_2)/2}K_{b_1-b_2}(2\sqrt{z})
\ee
and  
\be  \label{rep2}
G^{2,1}_{1,2}\left(\begin{array}{c} a_1 \\ {b_1,b_2} \end{array} , z \right) =
\Gamma(a_1+b_1+1)\Gamma(a_1+b_2+1)U(a_1+b_1+1,a_1+b_1+1,z)
\ee
where $U$ is the Kummer confluent hypergeometric function, related to the Whittaker function as
\be
W_{a,b}(z)=e^{-z/2}z^{b+1/2}U(b-a+1/2,1+2b,z).
\ee
These examples provide an illustration of the integration \eqref{conv}
\be  \label{example1}
\int_0^{\infty}G^{1,0}_{0,1}\left(\begin{array}{c}- \\ b_{1'} \end{array}, z \right)G^{2,0}_{0,2}\left(\begin{array}{c} - \\ {b_1,b_2} \end{array} , xz \right)dz = G^{2+0,0+1}_{0+1,2+0}\left(\begin{array}{c} -b_{1'} \\ {b_1,b_2} \end{array} , x \right) =  G^{2,1}_{1,2}\left(\begin{array}{c} -b_{1'} \\ {b_1,b_2} \end{array} , x \right)
\ee
which if we write it out using \eqref{rep1} and \eqref{rep2}  corresponds to
\be  \label{direct}
\int_0^{\infty}( z^{b_{1'}} e^{-z} )(2(xz)^{(b_1+b_2)/2}K_{b_1-b_2}(2\sqrt{xz}))\, dz = 
\Gamma(b_1+b_{1'}+1)\Gamma(b_{1'}+b_2+1)U(b_1+b_{1'}+1,b_1+b_{1'}+1,x)
\ee
provided $b_1+b_{1'}>-1$,  $b_2+b_{1'}>-1$, $(b_1+b_2)/2+b_{1'}>-1$. The result \eqref{direct} can be checked directly
using power series representations
of the Bessel $I_s(z)$ and the Kummer $M(a,b,z)$, that $K_s(z)$ and $U(a,b,z)$ are linear combinations of, respectively.

We can easily map example \eqref{example1} to similar formulas that might be less familiar.
For example,  using \eqref{inv} we can map between the two elementary examples $G^{0,1}_{1,0}$ and $G^{1,0}_{0,1}$ in \eqref{example1}:
\be    \label{invexample}
(z^{-1})^{b_1} e^{-z^{-1}} = 
G^{1,0}_{0,1}\!\left(\begin{array}{c} - \\ b_1 \end{array}, z^{-1} \right)
=G^{0,1}_{1,0}\!\left(\begin{array}{c} 1-b_1 \\ - \end{array}, z \right) = z^{(1-b_1)-1} e^{-1/z}
=z^{-b_1}e^{-1/z} \;. 
\ee
Note that we did not quite obtain $G^{0,1}_{1,0}(b_1, z^{-1})$, since
the arguments ${\bf a}_p$, ${\bf b}_q$ are  acted upon by the inversion \eqref{inv}. Mapping
$G^{1,0}_{0,1}$ in example \eqref{example1} to  $G^{0,1}_{1,0}$ we  find our second 
integration example
\be  \label{example2}
\int_0^{\infty}G^{0,1}_{1,0}\!\left(\begin{array}{c} a_{1'} \\ - \end{array}, z \right)G^{2,0}_{0,2}\!\left(\begin{array}{c} - \\ {b_1,b_2} \end{array} , xz \right)dz = G^{2+1,0+0}_{0+0,2+1}\!\left(\!\!\begin{array}{c} - \\ {-a_{1'},b_1,b_2} \end{array} ,x \right) =  G^{3,0}_{0,3}\!\left(\!\!\begin{array}{c} - \\ {-a_{1'},b_1,b_2} \end{array} , x \right)
\ee
which if we write it out using \eqref{rep1} and \eqref{rep2}  corresponds to
\be
\int_0^{\infty}( z^{a_{1'}-1} e^{-1/z} )(2(xz)^{(b_1+b_2)/2}K_{b_1-b_2}(2\sqrt{xz}))\, dz = 
 G^{3,0}_{0,3}\!\left(\!\!\begin{array}{c} - \\ {-a_{1'},b_1,b_2} \end{array} , x \right)
\ee
which unlike  \eqref{example1} is not a single hypergeometric function, since neither $m$ nor $n$ is equal to one
(cf. \eqref{hyperG} below). To check \eqref{example2}, we can compare to 3.16.3.9 in \cite{Prudnikov}, Vol.4.

Note that the difference between \eqref{example1} and \eqref{example2} is not
merely a change of variable of integration $z$: the inversion \eqref{invexample} only acted on the first factor
in the integrand, whereas a change of variables of course acts on both factors,  
and indeed the right hand sides of  \eqref{example1} and \eqref{example2} are not the same. The replacement of $z^{b_{1'}}$
by $z^{a_{1'}-1}$ 
can be absorbed by relating the parameters, but replacing $e^{-z}$ with $e^{-1/z}$  changes the analytic properties
of the integrand,
making it more like a combination of ${}_0F_2$ hypergeometric functions, whereas $U$ is a combination of
of ${}_1F_1$ hypergeometric
functions. This inversion in the exponent can come from S-transforming
a lattice theta function, so in this sense Meijer $G^{3,0}_{0,3}$ is ``dual'' to Kummer $U$.

One comment about representation: we see from definition \eqref{def1} that a power can always be absorbed in
a shift of all arguments
\be
z^{\mu}G^{m,n}_{p,q}\!\left(\begin{array}{c}{\bf a}_p \\ {\bf b}_q \end{array},z\right)
=G^{m,n}_{p,q}\!\left(\begin{array}{c}{\bf a}_p+{\bf \mu} \\ {\bf b}_q+{\bf \mu} \end{array},z\right)
\ee
where ${\bf a}_p+{\bf \mu}$ means adding $\mu$ to all components of the vector ${\bf a}_p$. 
For example, in  the special case ${\bf b}_q=(0,0,-s)$ of the integral \eqref{example2} that appears
in the main text,
we have two alternative representations:
\be
G^{3,0}_{0,3}\!\left(\!\!\begin{array}{c} - \\ {-s,0,0} \end{array} ,z \right)
=
z^{-s/2}G^{3,0}_{0,3}\!\left(\!\!\begin{array}{c} - \\ {-{s \over 2},{s\over 2},{s \over 2}} \end{array} ,z \right) \;. 
\ee

To end this appendix, we can compare the definition \eqref{def1} with the hypergeometric function ${}_pF_q$, that always has a power series representation. 
The Meijer $G^{m,n}_{p,q}$ does not necessarily have a single power series representation, but neither
does $K_s$ or $U$ --- each is a sum of two terms. Meijer $G^{m,n}_{p,q}$ has some other useful properties
as discussed above. Meijer G and hypergeometric functions are put on the same footing by the Barnes representation of hypergeometric functions,
where  Meijer $G^{m,n}_{p,q}$ is specialized to one of the upper arguments $m$ and $n$ being one,
say $m=1$, and $n=p$:
\be  \label{hyperG}
{}_p F_q\!\left(\!\!\begin{array}{c} {\bf a}_p \\ {\bf b}_q \end{array} , z \right)
={\Gamma({\bf b}_q) \over \Gamma({\bf a}_p)}
G^{1,p}_{p,q+1}\!\left(\!\!\begin{array}{c} 1- {\bf a}_p \\ {0, 1-{\bf b}_q} \end{array} ,-z \right) \; , 
\ee
and  $G^{2,1}_{1,2}$ in \eqref{example1} is related to $G^{1,2}_{2,1}$ (which is in the form \eqref{hyperG}
for $p=2$, $q=0$, i.e. ${}_0F_2$) by the inversion \eqref{inv}, so the argument is inverted
to $-z^{-1}$. 

A perhaps more familar example of \eqref{hyperG} is $p=2$, $q=1$ which gives ${}_2F_1$ in the Barnes representation:
\be
{}_2F_1(a,b,c;z) = {\Gamma(c) \over \Gamma(a)\Gamma(b)}{1 \over 2\pi i}
\int_{-i\infty}^{i \infty} {\Gamma(a+s)\Gamma(b+s)\Gamma(-s) \over \Gamma(c+s)}(-z)^s ds \; . 
\ee
Now, other Meijer G functions like $G^{3,0}_{0,3}$, where neither $m$ nor $n$ are equal to one
as in \eqref{hyperG},
are not hypergeometric, but they can be sums of hypergeometric functions. 
To make the connection to the main text, we note
that if the integral \eqref{def1} converges, and the $m$ factors $\Gamma(b_j-s)$ have no confluent poles, 
evaluation of  $G^{m,n}_{p,q}$ by residues gives $m$ terms with hypergeometric functions.
For  $G^{3,0}_{0,3}$  this means that if the differences between the arguments $(b_1,b_2,b_3)$ are not integers,
 $ G^{3,0}_{0,3}$ 
has a representation as a sum of $m=3$ terms with ${}_0\widetilde{F}_2$ hypergeometric functions\footnote{See
\href{https://functions.wolfram.com/HypergeometricFunctions/MeijerG/03/01/05/01}{Wolfram Functions,
Meijer G}.}:
\begin{eqnarray} \label{hyper}
 G^{3,0}_{0,3}\!\left(\!\!\begin{array}{c} - \\ {b_1,b_2,b_3} \end{array} , x \right)
 &=&\pi^2\left(
 {z^{b_1} {}_0\widetilde{F}_2(b_1-b_2+1,b_1-b_3+1,-z)  \over \sin(\pi(b_2-b_1))\sin(\pi(b_3-b_1))}
+\right.\\
&&\left. \hspace{-4cm}{z^{b_2} {}_0\widetilde{F}_2(1-b_1+b_2,b_2-b_3+1,-z) \over \sin(\pi(b_1-b_2))\sin(\pi(b_3-b_2))}
 + {z^{b_3}{}_0\widetilde{F}_2(1-b_1+b_3,1-b_2+b_3,-z) \over \sin(\pi(b_1-b_3))\sin(\pi(b_2-b_3))}\nonumber
  \right)
\end{eqnarray}
where the standard regularized hypergeometric function is
\be \label{Freg}
{}_0\widetilde{F}_2(b_1,b_2,z)= {{}_0F_2(b_1,b_2,z) \over \Gamma(b_1)\Gamma(b_2)}
\ee
that is finite for finite values of its arguments. Conversely, ${}_0F_2$ itself
has poles for example if an argument becomes zero, that is divided out in \eqref{Freg}.

In our example $(b_1,b_2,b_3)=(0,0,-s)$, to use the specialized expression \eqref{hyper} 
we would have to introduce regularization by hand,
but we emphasize that this is not in the spirit of the definition \eqref{def1} where the contour avoids poles
by definition. Still, it might be of some use to state what regularization of \eqref{hyper}  by 
hand would mean. For example, we can set $b_2=\delta> 0$
to prevent the first two denominators in \eqref{hyper} from  vanishing as $b_1-b_2$. 
Additionally we are interested in $s=0$ and $s=1$, and set $s=\epsilon$ or $s=1+\epsilon$ with $\epsilon>0$, respectively, and then set $\delta=\epsilon$, we have simple and double poles in $\epsilon$.
The choices $b_2=\delta$ and $\delta=\epsilon$ represent regularization ambiguities. However for our purposes we will not need \eqref{hyper},
we only provided it for context.

A final comment: one might ordinarily think of a Bessel function with a square root in its argument  $K_s(2\sqrt{x})$ as more complicated
than one without a square root, but to get a Bessel function {\it without} a square root we need to go one
step {\it higher} in the Meijer-G hierarchy, from $p=0$ in \eqref{rep2} to $p=1$:
\be  \label{Besselnos}
G^{2,0}_{1,2}\left(\begin{array}{c} 1/2 \\ {s,-s} \end{array} , z \right) = \pi^{-1/2}e^{-x/2}K_s(x/2) \;. 
\ee
So the Bessel function with a square root in the argument $K_s(2\sqrt{x})$, as we get from the massive Eisenstein series, is more basic than  $K_s(x/2)$ in the sense that
the former requires one less gamma function in the Mellin-Barnes expression \eqref{def1}. The $I$ Bessel function without a square root in the argument
is similar to \eqref{Besselnos} with $m=2,n=0$ replaced by $m=1,n=1$:
\be
G^{1,1}_{1,2}\!\left(\!\!\begin{array}{c} 1/2 \\ {s,-s} \end{array} ,z \right) \;  = \pi^{1/2}e^{-x/2}I_s(x/2) \; . 
\ee
This representation and the algorithmic convolution in \eqref{conv} could perhaps be of use
to combine the massive Eisenstein series with Niebur-Poincar\'e series.

\section{Three normalizations of $SL(2,{\mathbb Z})$ Eisenstein series}
\label{norm}
We
use the normalization from the book \cite{Fleig:2015vky}, Ch.\ 10 (Ch.\ 11 in the arXiv version),
where the seed is just $\tau^s$ with no additional factor. In particular,
\begin{equation} \label{Esbook}
E_{s} = \tau_2^s + {\xi(2s-1) \over \xi(2s)}\tau_2^{1-s}
+{2 \over \xi(2s)}\tau_2^{1/2}\sum_{m\neq 0}|m|^{s-1/2}\sigma_{1-2s}(m)K_{s-1/2}(2\pi|m|\tau_2)e^{2\pi i \tau_1}
\end{equation}
with the completed zeta function $\xi(s)=\pi^{-s/2}\Gamma(s/2)\zeta(s)$.
The pole at $s=1$ comes from the second term (constant $\tau_2^{1-s}=\tau_2^0$ in the limit $s=1$) and the residue  is $3/\pi$. 

We are typically interested in the first subleading term near a pole.
In the same reference \cite{Fleig:2015vky} section 10.1,
\be   \label{norm1}
E_s(\tau) = {3 \over \pi(s-1)} + {6 \over \pi}\left(12 \log A -\log(4\pi) - \log(\sqrt{\tau_2}|\eta(\tau)|^2)\right) + \ldots
\ee
where $A$ is the Glaisher constant $A=e^{1/12-\zeta'(1)}$. 

If we change overall normalization, there is both a multiplicative factor and an additive constant
in the subleading term. Explicitly, if the new normalization $N(s)$ is finite at $s=1$ (as will be the case in the examples below):
\begin{eqnarray} \label{shiftN}
N(s)E_s(\tau) &=&  (N(1) + N'(1)(s-1) + \ldots)\left( {3 \over \pi(s-1)} + E'_s(\tau)\Big|_{s=1} + \ldots\right) \nonumber \\
&=&N(1)\left(  {3 \over \pi(s-1)}  +E'_s(\tau)\Big|_{s=1}\right)  + {3 \over \pi} \cdot N'(1)  \; . 
\end{eqnarray}

The first common example of  normalization
other than \eqref{Esbook} is if the lattice sum is not restricted to mutually prime
$m,n$, that gives a change of normalization by $N=2\zeta(2s)$. We have from \eqref{shiftN} that
\be   \label{norm2}
2\zeta(2s)E_s(\tau) = {\pi \over (s-1)} + 2\pi \left(\gamma -\log(2) - \log(\sqrt{\tau_2}|\eta(\tau)|^2)\right)+ \ldots
\ee
where $\gamma$ is the Euler-Mascheroni constant.
The factor in front of the subleading term is $N(1)\cdot (6/\pi)=2\zeta(2)\cdot (6/\pi)=2\pi$, the shift is ${N'(1) \over N(1)} = {2 \zeta'(2) \over \zeta(2)} $
and \eqref{norm1} and \eqref{norm2} are consistent because
\be
12 \log A -\log(4\pi) = \gamma -\log(2) - { \zeta'(2) \over \zeta(2)} \; . 
\ee

A third common normalization is the completed version:
\begin{equation}
E^{\star}_s =\xi(2s)E_{s} = \pi^{-s}\Gamma(s)  {\zeta(2s)}E_{s}
\end{equation}
with the completed zeta function $\xi$ given at the beginning of this section. In particular
$\xi(2)=\pi^{-1}\Gamma(1)\zeta(2)=\pi^{-1}\pi^2/6=\pi/6$. 
(Note $\xi(2)/\zeta(2)=\pi^{-1}$.)

For $E^{\star}_s$,  we can check from \eqref{Esbook} that
the coefficient of $\tau_2^{1-s}$
is
\begin{equation}
\pi^{-s}\Gamma(s)\cdot \zeta(2s)\cdot {\xi(2s-1) \over \xi(2s)}
=\pi^{1/2-s}\Gamma(s-1/2)\zeta(2s-1)= {1 \over 2(s-1)} + \ldots
\end{equation}
so $E^*_s$  has residue 1/2 at $s=1$.

\section{Ramanujan duality}

If we know how an automorphic form transforms under Weyl reflection,
this will imply some summation identity for the Fourier sum. 
For example, if we Fourier expand the identity ${E}_s(\tau_2)=E_s(1/\tau_2)$
we find an identity that according to \cite{Cohen} was given by Ramanujan:
\begin{eqnarray} \label{raman}
&& \hspace{-2cm} 4\sqrt{x}\sum_{n=1}^{\infty}{\sigma_s(n) \over n^{s/2}}K_{s/2}(2\pi n x)=  
{4 \over \sqrt{x}}\sum_{n=1}^{\infty}{\sigma_s(n) \over n^{s/2}}K_{s/2}\left({2\pi n \over x}\right)\\[0mm]
 &&+\xi(-s)(x^{(1+s)/2}-x^{(1+s)/2})-  \xi(s)(x^{(1-s)/2}-x^{(s-1)/2}) \nonumber
\end{eqnarray}
which for $s=1$, when the Bessel functions reduce
to exponential functions, implies the modular transformation of $|\eta(\tau)|$.
We will refer to the summation \eqref{raman} as ``Ramanujan duality'' $x \leftrightarrow 1/x$.

In \cite{Berndt}, Ramanujan duality for Dirichlet series and $L$-functions is studied,
and we find a similar duality $r\leftrightarrow c$ as follows:
\begin{eqnarray} \label{B1}
\sum_{n=0}^{\infty}{\sigma_k(n) \over (c^2+n)^{\nu/2}}
K_{\nu}(4\pi r \sqrt{c^2+n})
&=&-{\delta_{1,k} \over 4\pi rc^{\nu-1}}K_{\nu-1}(4\pi r c)\\[-1mm]
&&+ {1 \over r^{\nu}c^{\nu-k-1}}\sum_{n=0}^{\infty}
{(-1)^{(k+1)/2}\sigma_k(n) \over (r^2+n)^{(k+1-\nu)/2}}K_{k+1-\nu}(4\pi c\sqrt{r^2+n}) \; . \nonumber
\end{eqnarray}
Another is for $\chi$ a primitive character modulo $q$:
\begin{eqnarray} \label{B2}
&&\hspace{-2cm}\sum_{n=0}^{\infty}{n \chi(n) \over (c^2+n^2/(2q))^{\nu/2}}
K_{\nu}(4\pi r \sqrt{c^2+n^2/(2q)})\\
&=& -{i\tau(\chi) \over r^{\nu}c^{\nu-3/2}\sqrt{q}}\sum_{n=0}^{\infty}
{n\bar{\chi}(n) \over (r^2+n^2/(2q))^{(3/2-\nu)/2}}K_{3/2-\nu}(4\pi c\sqrt{r^2+n^2/(2q)}) \; . \nonumber
\end{eqnarray}
Both \eqref{B1} and \eqref{B2} are reminiscient of the summation identity that follows from equating Fourier coefficients
for ${\mathcal E}_s(\tau_2)$ and ${\mathcal E}_s(1/\tau_2)$. It is natural to expect that this equivalence would yield similar summation identities as in \eqref{B1} and \eqref{B2}, but we leave the details of this for future work.

\end{document}